\documentclass{article}

\usepackage{graphicx,bm,url,microtype,color}
\usepackage{amsfonts,amssymb,amsthm,amsmath}
\usepackage[english]{babel}
\usepackage[a4paper,text={16cm,23.5cm},centering]{geometry}
\usepackage[compact,small]{titlesec}
\usepackage[utf8]{inputenc}
\newcommand{\bx}{\mathbf x}

\newcommand{\bH}{\mathbf H}
\newcommand{\bI}{\mathbf I}
\newcommand{\bX}{\mathbf X}

\newcommand{\bbR}{\mathbb R}
\newcommand{\PPA}{\mbox{PPA}}
\newcommand{\TSD}{\mbox{TSD}}

\newenvironment{keywords}
    {\vspace*{3mm}
    {\noindent{}\textit{Keywords\/:}}
        \nopagebreak\small}
        {}

\newenvironment{MSC}
    {\vspace*{3mm}
    {\noindent{}\textit{MSC\/:}}
        \nopagebreak\small}
        {}

\title{A survey and a new selection criterion\\ for statistical home range estimation}
\author{A. Ba\'{\i}llo$^a$ and J.E. Chac\'{o}n$^b$ \\
{\normalsize $^a$ Departamento de Matem\'{a}ticas, Universidad Aut\'{o}noma de Madrid (Spain)} \\
{\normalsize $^b$ Departamento de Matem\'{a}ticas, Universidad de Extremadura (Spain)} }

\begin{document}

\maketitle

%\color{black}

%\color{rojo_oscuro}
\begin{abstract}
\noindent
The home range of a specific animal describes the geographic area where this individual spends most of the time while carrying out its usual activities (eating, resting, reproduction, \ldots). Although a well-established definition of this concept is lacking, there is a variety of home range estimators. The first objective of this work is to review and categorize the statistical methodologies proposed in the literature to approximate the home range of an animal, based on a sample of observed locations. The second aim is to address the open question of choosing the ``best'' home range from a collection of them based on the same sample. We introduce a numerical index, based on a penalization criterion, to rank the estimated home ranges. The key idea is to balance the excess area covered by the estimator (with respect to the original sample) and a shape descriptor measuring the over-adjustment of the home range to the data. To our knowledge, apart from computing the home range area, our ranking procedure is the first one which is both applicable to real data and to any type of home range estimator. Further, the optimization of the selection index provides in fact a way to select the smoothing parameter for the kernel home range estimator. For clarity of exposition, we have applied all the estimation procedures and our selection proposal to a set of real locations of a Mongolian wolf using R as the statistical software. As a byproduct, this review contains a thorough revision of the implementation of home range estimators in the R language.
\end{abstract}
%\color{black}
%------------------
\begin{keywords}
home range; utilization distribution; nonparametric; set estimation; selection criterion; telemetry; penalized criterion; animal movement; kernel density estimation
\end{keywords}

%-----------------
\begin{MSC}
Primary 62P12; secondary 62-02, 62G99.
\end{MSC}

\newpage

\section{Introduction to the home range problem} \label{Section.Introduction}

\subsection{Problem statement} \label{Subsection.ProblemStatement}

There has long existed an interest in identifying different characteristics (degrees of use, geographical limits, environmental descriptors...) of space use by whole animal species or {\em individuals}.
The space use can be described by various (sometimes interrelated) concepts. In this work we specifically focus on the concept of {\em home range}.
As Seton (1909, p. 26) pointed out when speaking about the home range of an individual, ``no wild animal roams at random over the country; each has a home-region, even if it has not an actual home. The size of this home region corresponds somewhat with the size of the animal. [\ldots] In the idea of a home-region is the germ of territorial rights.''.

Burt (1943) is credited with the first formalization of the idea of home range, as ``that area traversed by the individual in its normal activities of food gathering, mating and caring for young. Occasional sallies outside the area, perhaps exploratory in nature, should not be considered as in part of the home range.'' Resting and watering areas and the routes traveled between all these strategic locations may also be part of the home range. The home range of an animal should not be confused with its territory, which is the defended portion of the home range. In particular, home ranges of different individuals may overlap (for instance, water sources can be shared).

In everyday practice, conservation biologists and spatial ecologists, among others, have a keen interest in estimating space use maps from animal tracking data, for example for monitoring threatened species and for conservation planning.
Further, the widespread use of geolocated smartphones has lead to massive amounts of analogous tracking data for billions of human individuals (Meekan {\em et al.} 2017). Analysis of space use is, thus more appealing than ever, not only from a conservational point of view (analysis of interactions between animals and humans), but also from an economical or even anthropological perspective (Walsh {\em et al.} 2010).

Delineation of the home range boundary is a popular and simple way of describing the space use of a monitored individual. This problem has been tackled in a variety of ways of increasing complexity.
For instance, individual space use can also be described by the {\em utilization distribution} (UD), the probability distribution of the animal locations over a period of time (van Winkle 1975). Nowadays, {\em statistical} procedures aiming to obtain the home range usually define it as a level set, with high probability content (e.g., 95\%), of the utilization density, $f$.
Another appealing possibility is to use {\em mechanistic} home range analysis (see Moorcroft and Lewis 2006), which first proceeds by modeling the trajectory of the animal via a stochastic process.

The purpose of this work is twofold. First we provide a thorough review of the statistical procedures proposed in the literature to estimate the home range of a specific animal.
At the same time, along with the conceptual review, the libraries in the statistical software R (R Core Team, 2018) available for the different procedures will be also surveyed.
%Mostly, these techniques, of a descriptive nature, are set estimation ones, either through direct geometrical procedures to estimate a set or through density estimation of the UD and posterior obtention of its level set.
The second aim of this work stems precisely from this review. Given such a wealth of possible estimators, which is the ``best'' one? To our knowledge, this is an interesting and still open question. In Section~\ref{Section.Selection} we propose to construct a numerical index that balances overestimation and overfitting of the home range with respect to the observed locations. An advantage of our proposal is that it can be computed for any type of home range estimator, while previously existing selection procedures only work for specific types of home ranges (for example, those defined as the level set of a utilization density).

\subsection{Characteristics of animal location data} \label{Subsection.AnimalLocations}

Originally, animal locations were obtained using live traps (Blair 1940) or continuous observation of the animal (Odum and Kuenzler 1955). In the 1960s these techniques were substituted by radio telemetry, which consists in attaching a transmitter to the animal and recording its position via the transmission of radio signals.
Recent advances in tracking and telemetry technology (e.g., GPS instead of radio emissions) provide animal location data with an even higher frequency and accuracy (Kie {\em et al.} 2010). This, together with improvement in computing resources, has originated the use of new estimation techniques and an increasing generalized interest in the analysis of space use based on tracked individuals.

Fleming {\em et al.} (2015) highlight the fact that animal tracking data are, by nature, autocorrelated, an obvious consequence of animal movement being a continuous process. Consequently, these authors define a home range as a region with a prespecified probability content (usually 95\%) of the {\em range distribution}, which is the ``probability distribution of all possible locations as determined from the distribution of all possible paths''. Fleming {\em et al.} (2015) consider that an autocorrelated sample contains less geometrical information about the density level set contour than a sample of independent observations of the same size. However, the existence of autocorrelation between observations implies that we can take advantage of information from the past to predict the future movements of the animal and, thus, it should be used for home range estimation.

\subsection{A real data set: Mongolian wolves} \label{Subsection.MongolianWolves}

\begin{figure}[h]
\begin{center}
\includegraphics[width=15cm]{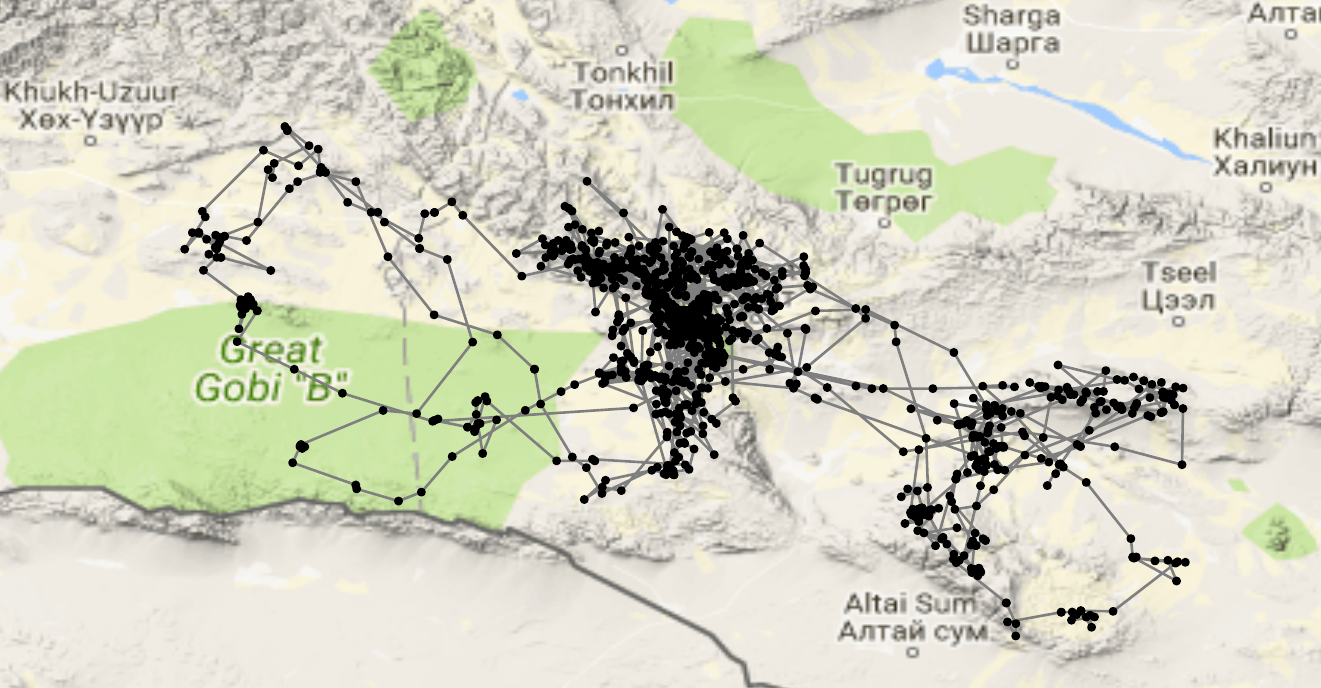}
\end{center}
\caption{Relocations and trajectories of the Mongolian wolf Zimzik on Google Maps.}
\label{Figure.MongolianWolvesMovebank}
\end{figure}

Throughout this work we will be reviewing techniques for home range estimation proposed in the literature. The procedures will be illustrated via the analysis of the real data set of relocations with ID 14291019 downloaded from Movebank (\url{www.movebank.org}), an animal movement database coordinated by the Max Planck Institute for Ornithology, the North Carolina Museum of Natural Sciences, and the University of Konstanz. This data set contains the locations of a pair of Mongolian wolves, Boroo (female) and Zimzik (male), observed during a period starting May 2003 (Boroo) or March 2004 (Zimzik) and ending September 2005 (both) with GPS technology (Kaczensky {\em et al.} 2006, Kaczensky {\em et al.} 2008). Signal transmission took place between one to three times a day, at irregularly-spaced times. Between the two wolves we chose Zimzik, {since} its trajectory, with 1455 observed locations {as depicted in the map} in Figure~\ref{Figure.MongolianWolvesMovebank}, seems to correspond to a home range with interesting mathematical features (more than one connected component, non-convex components, \ldots). The map comprises part of the provinces of Hovd and Govi-Altay in southwestern Mongolia (the provinces {border is shown as a dashed grey line}) and China (the Mongolia-China border is in solid grey). We can also see the patches of the Great Gobi B Strictly Protected Area, an International Biosphere Reserve in Gobi desert, and other nature reserves. As noted by Kaczensky {\em et al.} (2008), in Figure~\ref{Figure.MongolianWolvesMovebank} we can see that Zimzik preferred mountainous terrains (slope $>5^{\circ}$), which hinder wolf hunting, over the flat steppe. Consequently, in this case elevation contours contribute to the shape of the home range. In general, it seems reasonable to believe that there are usually explanatory variables related (to some extent) to the utilization distribution and the home range of an animal. The problem of how to incorporate this information onto the home range estimator has been little analyzed yet (see Horne {\em et al.} 2008).

\section{Statistical analysis of positional data} \label{Section.SurveyEstimationHomeRange}

The statistical procedures to estimate home ranges are of a descriptive and nonparametric nature. They either estimate the home range directly, using geometric-type procedures, or they estimate first the utilization density and then compute the 95\% level set of the density estimator. The earliest estimation techniques were based on scarce location data, which were assumed to be independent. However, the high-frequency observation in current tracking samples demands incorporating time dependence in the analysis of positional data. To distinguish between these two setups, in Subsection~\ref{Subsection.IndependentLocations} we review the simpler home range estimators proposed under the independence assumption, and in Subsection~\ref{Subsection.TimeDependentHR} we introduce the more recent proposals dealing with time-dependent locations.

\subsection{Assuming location data are independent} \label{Subsection.IndependentLocations}

In this subsection we denote the animal relocations by $\bx_1,\ldots,\bx_n$ and we assume they are independent realizations of a random vector $\bX$ taking values in $\bbR^2$. In this setup, we begin by introducing global methods for home range estimation. The term ``global'' refers to the fact that each of such procedures is applied to the whole data set at once. In contrast, later we focus on the so-called localized methods, which stem from the former by applying a particular global method to each of the subsets of a suitably chosen partition of the data and then gathering up all the local pieces thus obtained. %\vspace{2 mm}

\subsubsection{Global methods} \label{Subsubsection.GlobalMethods}

\noindent
\emph{Minimum convex polygon (MCP) or convex hull} \vspace{2 mm}

The convex hull of a sample of points is the minimum convex set enclosing them all, yielding a polygon connecting the outermost points in the sample and all whose inner angles are less than 180$^{\circ}$. This is the simplest method for constructing home ranges and computing their areas and it has been widely employed for a long time, even until recently.

A variant of the MCP home range estimator is obtained by removing the proportion $\alpha$ of the sample points farthest from the sample centroid. More generally, Bath {\em et al.} (2006) compare the MCP home range after applying different criteria to ``peel'' individual observations from the sample of locations: they remove points farthest from the centroid or from the harmonic mean or those with a great influence on the area of the home range.

References using the minimum convex polygon (alone or in comparison with other methods) to estimate the home range of an individual are very abundant. Let us just mention a few: Mohr (1947), Odum and Kuenzler (1955), Worton (1987), Carey {\em et al.} (1990), Harris {\em et al.} (1990), Pavey {\em et al.} (2003), List and Macdonald (2003), Nilsen {\em et al.} (2008), Signer {\em et al.} (2015).

\begin{figure}[h]
\begin{center}
\includegraphics[width=9cm]{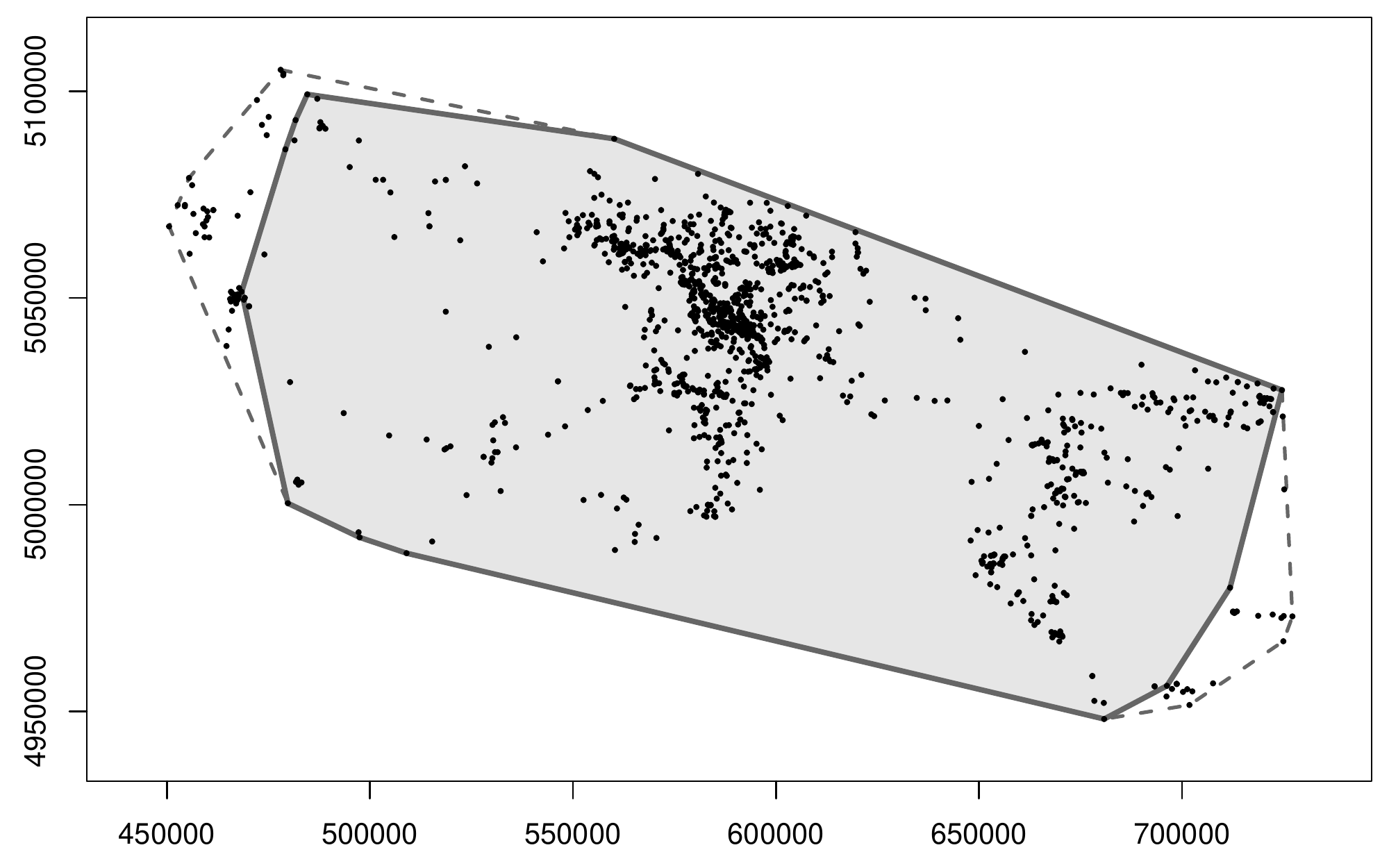}
\end{center}
\caption{Minimum convex polygons for Zimzik location data: the MCP containing 95\% of sample points is colored in grey. The dashed lines correspond to the MCP of the whole sample.}
\label{Figure.Wolves.MCP}
\end{figure}

Obviously, the convexity restriction has serious drawbacks, among them home range overestimation, as illustrated in Figure~\ref{Figure.Wolves.MCP}, showing the MCP of Zimzik locations. The MCP was computed using the R package {\tt adehabitatHR} (Calenge 2006). This home range does not adapt to the mountainous territory to which this wolf usually circunscribed its movements. To overcome this drawback, Harvey and Barbour (1965) proposed the minimum area method (also called by Kenward {\em et al.} 2001 the concave polygon). First, they computed the range length, the distance between the two locations furthest apart. Then, one outermost point (that is, in the boundary of the MCP) is connected with the next outermost one lying inside a ball of radius one quarter of the range length. The resulting minimum area home range is not necessarily convex and adapted better to the underlying sample shape.

\vspace{2 mm}

%\vspace{2 mm}

\noindent
\emph{$\alpha$-convex hull} \vspace{2 mm}

In the Computational Geometry literature, the notion of $\alpha$-convexity (first employed by Perkal 1956) is introduced as a generalized convexity concept, in order to relax the assumption of convexity, which can be very restrictive in practice. This is especially true in the problem of home range estimation, as illustrated above.

\begin{figure}[h]
\begin{center}
\includegraphics[width=9cm]{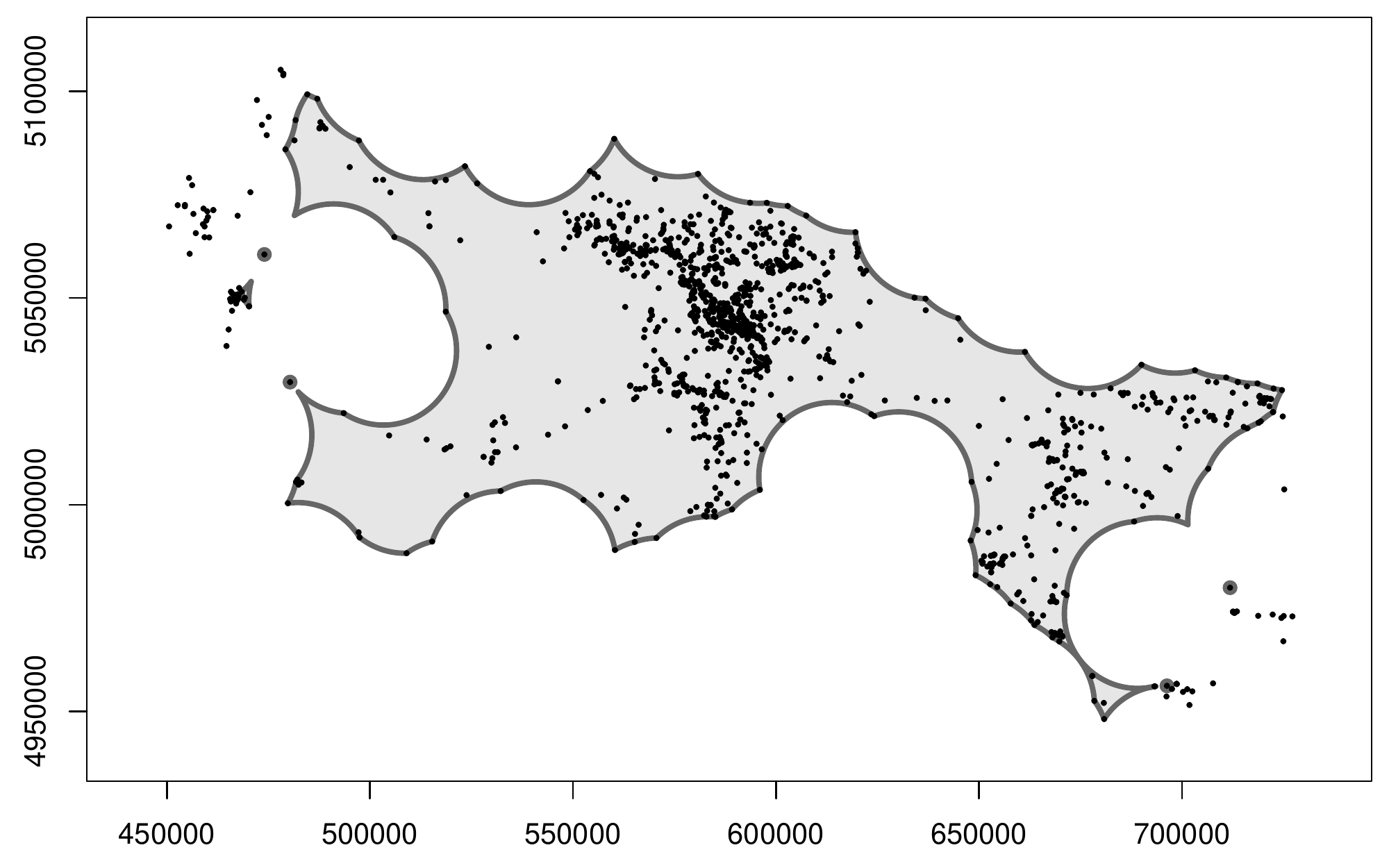}
\end{center}
\caption{$\alpha$-convex hull estimation of the home range for Zimzik locations.}
\label{Figure.Wolves.alphaCH}
\end{figure}

Rodr\'{\i}guez-Casal (2007) introduced the $\alpha$-convex hull of a sample of points and studied its convergence properties. Defined in simple terms, the $\alpha$-convex hull is the intersection of the complements of all the open balls of radius $\alpha$ which do not contain any sample points. Figure \ref{Figure.Wolves.alphaCH} shows the $\alpha$-convex hull of the observed Zimzik locations, with a value of $\alpha=18000$, computed using the R package {\tt alphahull} (Pateiro-L\'{o}pez and Rodr\'{\i}guez-Casal, 2016) after removing 5\% of the data points farthest from the data centroid (see also Pateiro-L\'{o}pez and Rodr\'{\i}guez-Casal, 2010).

The boundary of this home range estimator clearly shows the arcs of some of the balls of radius $\alpha$ that form the complement of the estimator. The improvement over the convex hull is clearly noticeable, since the $\alpha$-convex hull does not contain some of the large portions of never-visited terrain that the convex hull enclosed.

The radius $\alpha$ controls the regularity of the resulting estimator: large values of $\alpha$ make the $\alpha$-convex hull very similar to the ordinary convex hull, and small values yield a quite fragmented home range estimator. The suggested value of $\alpha=18000$ was chosen here by visual inspection, but a recent research study proposes an automatic, data-based choice of this parameter (see Rodr\'{\i}guez-Casal and Saavedra-Nieves, 2016).

Let us point out that there are other existing home-range estimation techniques that receive a similar name to ``$\alpha$-convex hull'', despite they refer to a completely different concept. To avoid confusion, it is worth mentioning the procedure of Burgman and Fox (2003), called $\alpha$-hull, which simply consists in obtaining the Delaunay triangulation of the data points and eliminating those line segments longer than a multiple $\alpha$ of the average line length. Although this latter $\alpha$-hull was initially introduced to estimate a species range, it has also been used for home range estimation (see, e.g., Foley {\em et al.} 2014).

\begin{figure}[h]
\begin{center}
\includegraphics[width=9cm]{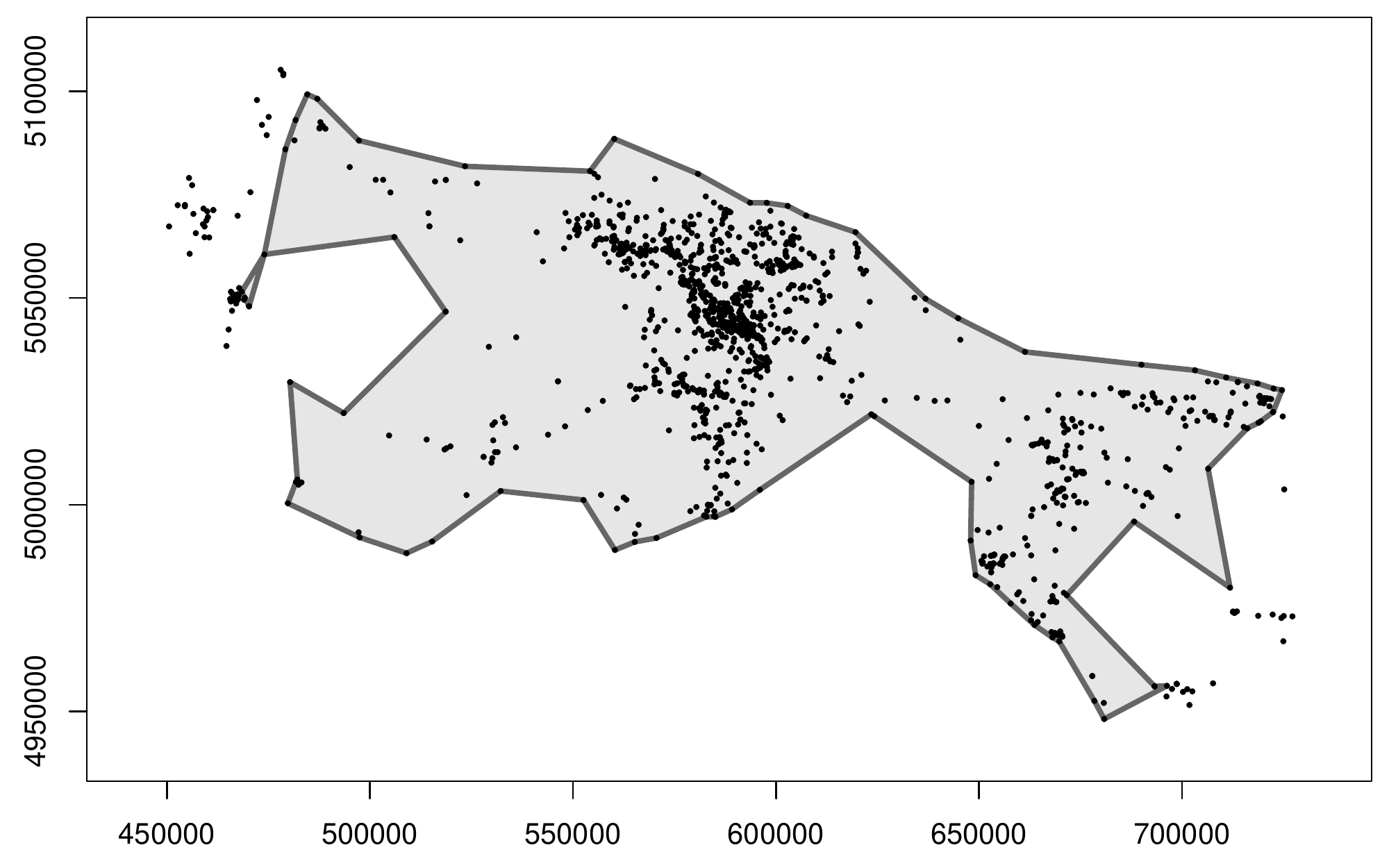}
\end{center}
\caption{$\alpha$-shape estimation of the home range for Zimzik locations.}
\label{Figure.Wolves.alphaS}
\end{figure}

In fact, Burgman and Fox's $\alpha$-hull is closer to another related object in the Computational Geometry literature: the $\alpha$-shape of the data points. This is another generalization of the convex hull, defined as the polygon whose edges are those segments connecting two data points if there is an open ball of radius $\alpha$ having those two data points in its boundary and no other data point in its interior (Edelsbrunner {\it et al.}, 1983). The $\alpha$-shape corresponding to Zimzik locations, computed with the R package {\tt alphahull}, is depicted in Figure \ref{Figure.Wolves.alphaS}, for the same value of $\alpha=18000$ as before and after removing the 5\% of the data points farthest from the data centroid. Its appearance is very similar to the $\alpha$-convex hull in Figure \ref{Figure.Wolves.alphaCH}, with the notable difference that its boundary is formed by straight lines, instead of circular arcs.

Still other recent generalizations of the convex hull are available for independent relocations, as for instance the cone-convex hull of Cholaquidis {\it et al.} (2014).
In the case of dependent locations, Cholaquidis {\em et al.} (2016) propose to estimate the home range by the $\alpha$-convex hull of the trajectory or by the reflected Brownian motion (RBM) sausage (the outer parallel set of the trajectory).
Here we will not explore these many variants. \vspace{2 mm}

\noindent
\emph{Harmonic mean home range} \vspace{2 mm}

\begin{figure}[h]
\begin{center}
\includegraphics[width=9cm]{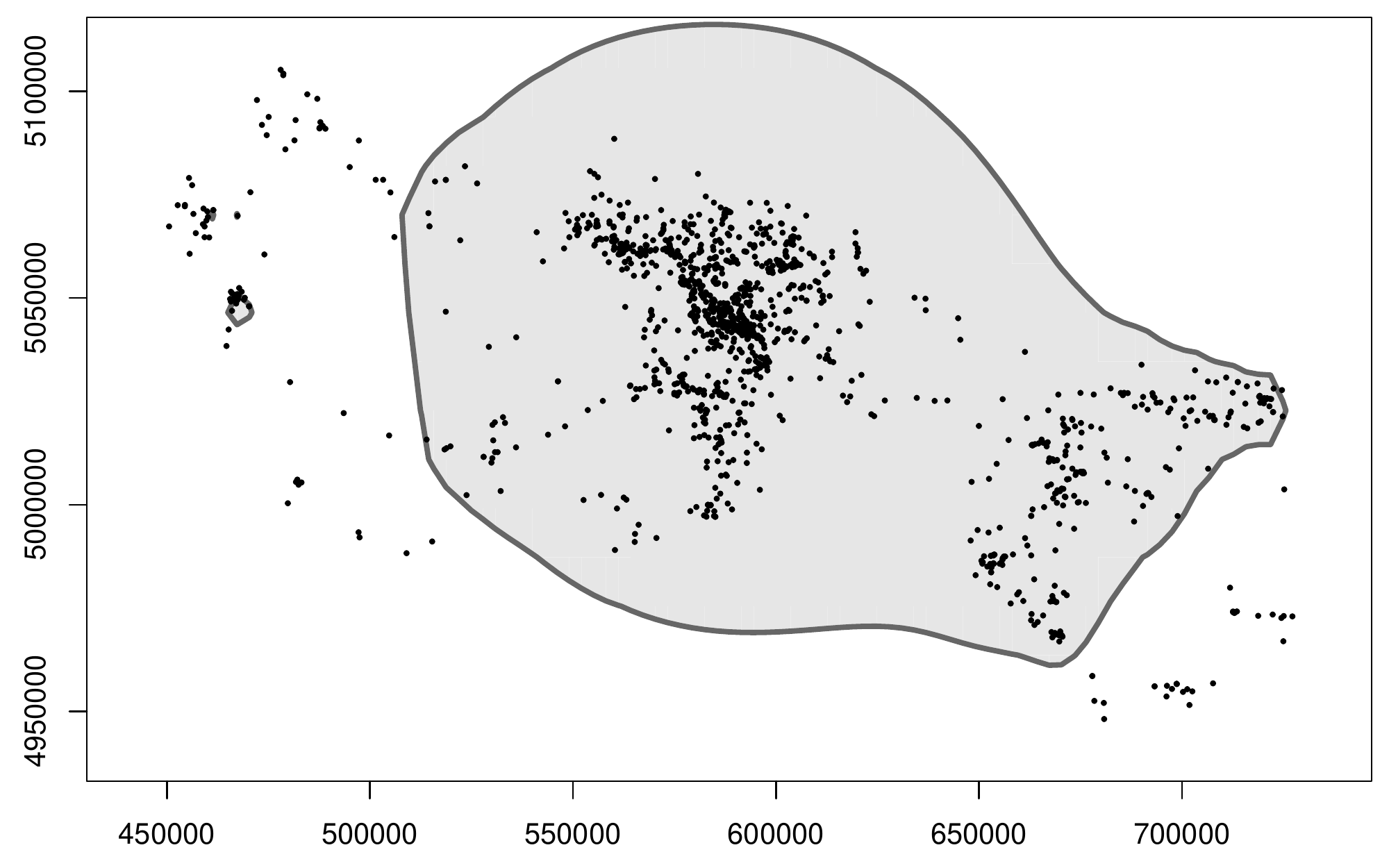}
\end{center}
\caption{Harmonic mean home range for the wolf data.}
\label{Figure.Wolves.HM}
\end{figure}

Dixon and Chapman (1980) define the areal sample moment of order -1 with respect to a point $\bx\in\bbR^2$ as the harmonic mean of the distances of $\bx$ to the observed locations:
\begin{equation*}
\mbox{HM}(\bx) = \frac{1}{n^{-1}\sum_{i=1}^n \|\bx-\bx_i\|^{-1}},
\end{equation*}
where $\|\bx-\bx_i\|$ is the Euclidean distance between $\bx$ and $\bx_i$. Then, as pointed out in Worton (1989), $\mbox{HM}(\bx)^{-1}$ can be considered as a kernel-type estimator of the UD, so that the harmonic mean home range is a level set of $\mbox{HM}^{-1}$ containing 95\% of the locations. In fact, Devroye and Krzy\.zak (1999) showed that  $\mbox{HM}(\bx)^{-1}$ is not a consistent estimator of the density, although this flaw is easily amended after a simple normalization, which consists of dividing $\mbox{HM}(\bx)^{-1}$ by $V_d\log n$, where $V_d$ is the volume of the unit ball in $\mathbb R^d$. In such form, it is called Hilbert kernel density estimate. In any case, note that the 95\% level set is unaffected by that normalization. We display in Figure~\ref{Figure.Wolves.HM} the resulting home range, which obviously shares many of the drawbacks of the MCP.
\vspace{2 mm}

\noindent
\emph{Kernel density estimation (KDE)} \vspace{2 mm}

Let us recall that $f$ denotes the utilization density. The home range of an animal is frequently defined as the level set $\{f\geq c\}$ of the utilization density attaining a 95\% probability content, that is, the level set associated to the level $c=c_{0.05}$ such that
$$
0.95 = \int_{\{f\geq c_{0.05}\}} f(\bx)\, d\bx.
$$
Thus, a plug-in estimator of the home range is the analogous level set $\{\hat f\geq \hat c_{0.05}\}$ of an estimator $\hat f$ of $f$, where
$$
0.95 = \int_{\{\hat f\geq \hat c_{0.05}\}} \hat f(\bx)\, d\bx.
$$
The first proposals along these lines assumed a bivariate normal model for the UD (Calhoun and Casby 1958). Still in a parametric setting, Don and Rennolls (1983) used a mixture of normal densities to estimate $f$. However, noting that such parametric models were usually inappropriate, in a seminal paper Worton (1989) introduced kernel density estimators as a natural nonparametric approach in home range procedures.

The general expression of a kernel density estimator is
\begin{equation} \label{KDE}
\hat f(\bx) = \frac{1}{n} \sum_{i=1}^n K_\bH (\bx-\bx_i),
\end{equation}
where $K:\bbR^2\to[0,\infty)$ is the kernel function (in this case, a probability density in $\bbR^2$), $\bH=(h_{ij})$ is a symmetric, positive-definite $2\times 2$ bandwidth matrix, and the scaling notation $K_\bH(\bx) := |\bH|^{-1/2} K(\bH^{-1/2}\bx)$ has been used, with $\bH^{-1/2}$ standing for the inverse of the matrix square root of $\bH$.

It is well known that the choice of the kernel $K$ has little effect on the accuracy of the estimator $\hat f$, compared to that of the bandwidth $\bH$. Worton (1989) chose a constrained bandwidth matrix $h^2\bI$ (where $\bI$ denotes the identity matrix)  depending on a single smoothing parameter $h>0$, which was proposed to be selected either via the ``ad hoc'' method (i.e., optimal for the Gaussian distribution) or via least-squares cross-validation. Worton (1989) additionally considered an adaptive KDE, that is, a kernel-type estimator where the bandwidth is allowed to vary from one observation $\bx_i$ to the other.

\begin{figure}[h]
\begin{center}
\begin{tabular}{cc}
\includegraphics[width=7.5cm]{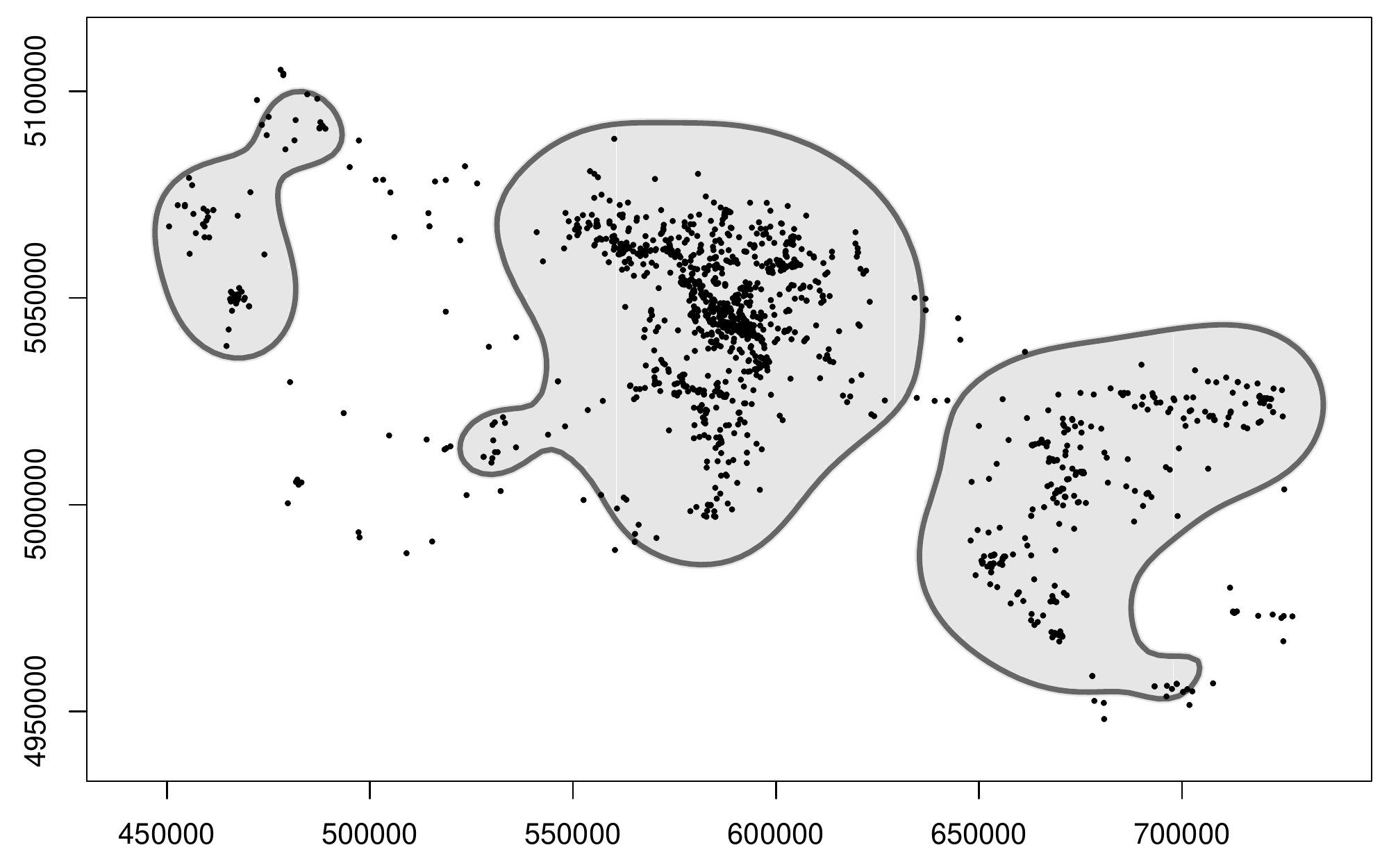} & \includegraphics[width=7.5cm]{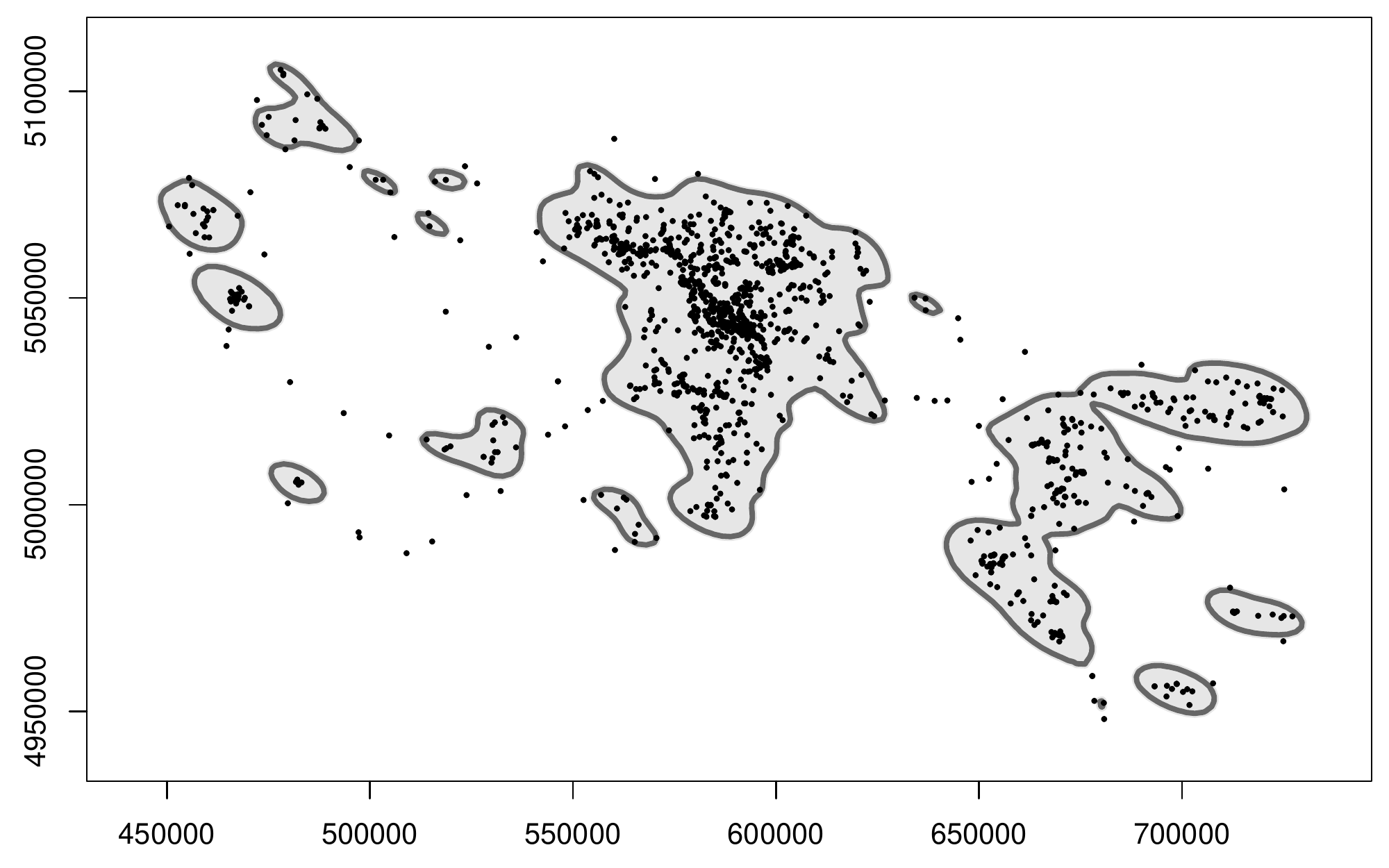} \\ (a) & (b)
\end{tabular}
\end{center}
\caption{KDE home ranges for Zimzik data using (a) the one-dimensional ad hoc bandwidth $h$ and (b) the unconstrained plug-in bandwidth matrix $\bH$.}
\label{Figure.KDE}
\end{figure}

The KDE home range for the wolf data is displayed in Figure~\ref{Figure.KDE} for the one-dimensional ad hoc bandwidth $h=12269.09$ (as computed using package {\tt adehabitatHR}) and the bidimensional unconstrained plug-in bandwidth matrix
\begin{equation} \label{UnconstrainedPIbandwidth}
\bH = \left( \begin{array}{rr}
23727441 & -9074807 \\
-9074807 & 10663700
\end{array} \right)
\end{equation}
of Chac\'{o}n and Duong (2010), obtained with the package {\tt ks} (Duong 2018). It is worth mentioning that, due to the very large values of the UTM coordinates, it was necessary to {pre-spherify the location data in order for the numerical routine to be able to perform the optimization correctly. This means that the data were pre-multiplied by ${\bf S}^{-1/2}$, the inverse of the square root of their sample variance matrix $\bf S$, so that the sample variance matrix of the transformed data became the identity matrix. A plug-in bandwidth was obtained from these sphered data and was then scaled back by pre- and post-multiplying it by the $\bf S$ to finally obtain the above plug-in matrix for the original data (see Duong and Hazelton, 2003, Section 4).}

As already noted by Bauder {\em et al.} (2015), the use of this unconstrained plug-in bandwidth matrix outperforms the single smoothing parameter approaches. The selected plug-in bandwidth matrix above suggests that, for this data set, it is advisable to use individual kernels with correlation coefficient $-0.571$, hence having their mass obliquely oriented, and not parallel to the coordinate axes as it happens when a diagonal or a single-parameter bandwidth is employed.

Moreover, these UD estimators clearly reveal that just trimming points furthest away from the data centroid to obtain a region containing 95\% of the data points might not be a good idea, since those points may not necessarily correspond to locations within low density zones.

Since its introduction, the use of the KDE home range has become widespread and a great deal of publications make use of it, even if the density estimator given in (\ref{KDE}) is originally thought for independent observations (see Subsection~\ref{Subsection.TimeDependentHR}). To cite a few case studies using this KDE home range, Bertrand {\em et al.} (1996), Kie and Boroski (1996), Girard {\em et al.} (2002), Hemson {\em et al.} (2005), Berger and Gese (2007), Pellerin {\em et al.} (2008), Jacques {\em et al.} (2009). Seaman and Powell (1996) and Seaman {\em et al.} (1999) analyze the KDE home range performance via simulations. Matthiopoulos (2003) presents a modification of the KDE to incorporate additional information on the animal.

Another ingenious modification of the KDE of a home range, namely the permissible home range estimator (PHRE), is given by Tarjan and Tinker (2016). These authors transform the original geographical sighting coordinates into coordinates with respect to relevant landscape features influencing animal space use. A KDE is constructed on the new coordinates and, afterwards, the corresponding estimated UD is backtransformed to geographical coordinates. The PHRE makes full sense in the context considered by Tarjan and Tinker (2016), but it is not clear that, in general, relevant features for the animal can always give rise to a new coordinate system. \vspace{2 mm}

\subsubsection{Localized methods} \label{Subsubsection.LocalizedMethods}

The previous global methods share the common feature that they employ the whole data set (or a slightly trimmed subsample) \emph{at once} to construct the home range estimate, either by applying some geometric construction to the sample points (convex hull and its variants) or by previously obtaining an estimate of the utilization density from them and then computing the density level set with 95\% probability content (kernel methods).

In contrast, localized methods proceed in three steps: first, a local neighbourhood of points is selected according to some criterion (pre-clustering, nearest neighbours, points {lying on a} certain ball); then, one of the previous global methods is applied only to the selected points to obtain a local home range estimate, and, finally, many of these local home ranges are gathered together by taking their union.
This way, many variants can be considered, depending on two choices: the way in which local neighbourhoods are constructed and the global method applied to each of these neighbourhoods. Next we describe the most popular ones.

\vspace{2 mm}

\noindent
\emph{Local convex hull (LoCoH) or k-nearest neighbor convex hulls (k-NNCH)} \vspace{2 mm}

This is a localized version of the MCP. For a fixed integer $k>0$, Getz and Wilmers (2004) construct the convex hull, $k$-NNCH$_i$, of each sample point $x_i$ and its $k-1$ nearest neighbors (NN) (with respect to the Euclidean metric, although other metrics could be employed). Then these hulls are ordered according to their area, from smallest to largest. The LoCoH home range estimate is the isopleth that results of progressively taking the union of the hulls from the smallest upwards, until a specific percentage (e.g., 95\%) of sample points is included.
Getz {\em et al.} (2007) extend the original LoCoH procedure (called $k$-LoCoH) to the $r$-LoCoH (where a fixed sphere of influence was used instead of the $k$ nearest neighbours) and to the $a$-LoCoH (the $a$ standing for adaptive sphere of influence). A detailed study of the theoretical properties and finite sample performance of the $r$-LoCoH can be found in Aaron and Bodart (2016).

The optimal choice of the number of neighbours, $k$, depends on the topological features of the home range. In particular, one possibility is to choose the minimal $k$ resulting in a prefixed genus (number of holes), as long as this information is known. If it is not, Getz and Wilmers (2004) suggest guessing the genus after inspection of the physical characteristics of the territory. Another possibility is to examine the areas of the isopleths as a function of the values of $k$ (see the documentation of the R package {\tt tlocoh}, Lyons {\em et al.} 2018).

The idea of incorporating topological information on the estimator has been little explored in the home range estimation literature. However, it represents a very active area in statistical research recently, encompassing a variety of methodologies under the name of Topological Data Analysis, which are nicely reviewed in Wasserman (2018) and implemented in the R package {\tt TDA} (Fasy {\it et al.}, 2017). Undoubtedly, the application of these modern techniques to the field of home range estimation shows promise as a very interesting venue for further research.

\begin{figure}[h]
\begin{center}
\includegraphics[width=9cm]{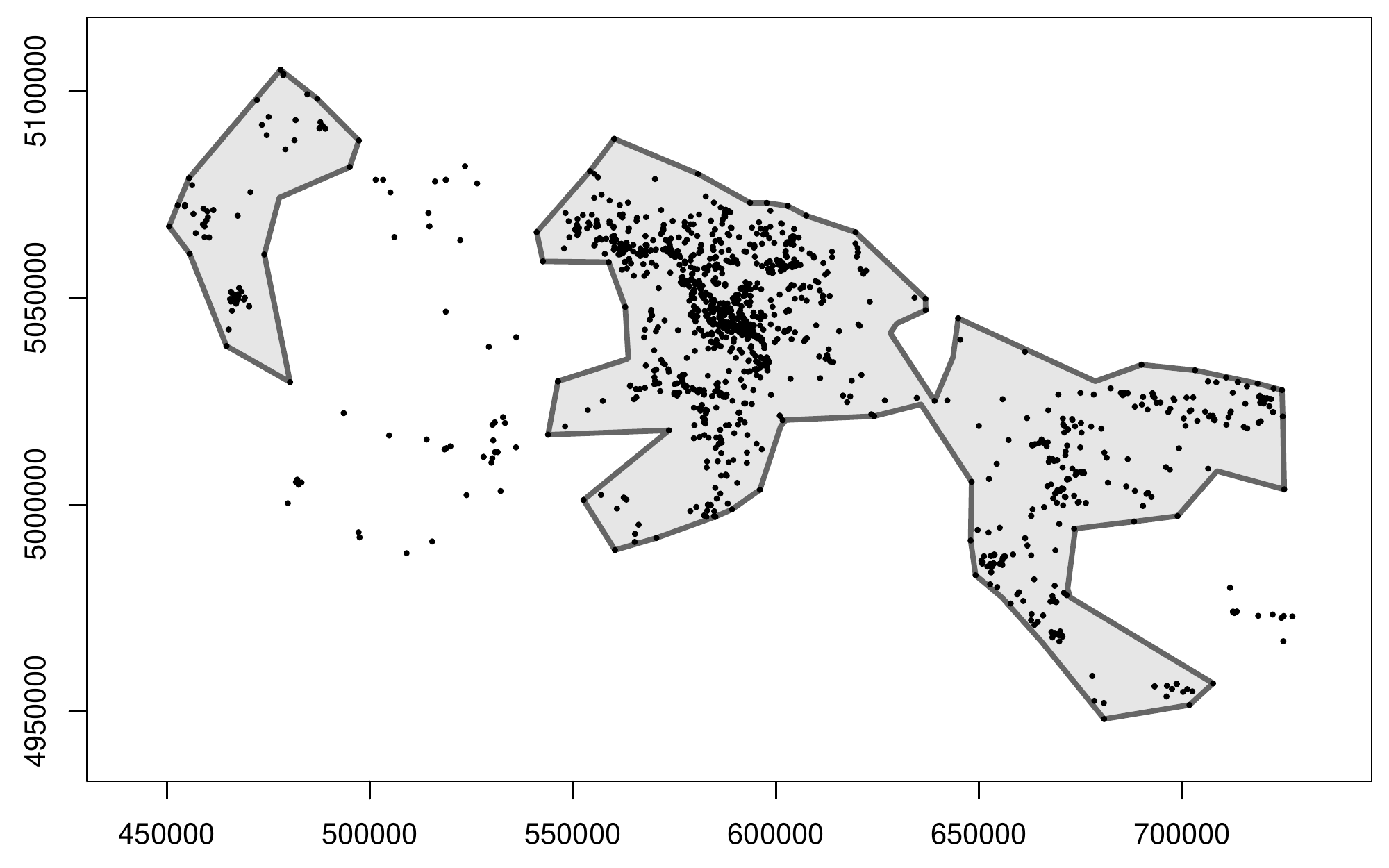}
\end{center}
\caption{LoCoH isopleths with probability content 95\% and $k=35$ neighbours for Zimzik locations.}
\label{Figure.Wolves.LoCoH}
\end{figure}

%The LoCoH procedure is implemented in the package {\tt adehabitatHR}, but it gives problems. Here I have used the package {\tt tlocoh}.
In Figure~\ref{Figure.Wolves.LoCoH} we display the LoCoH home range for Zimzik data and $k=35$ neighbours, obtained with {\tt tlocoh}. It is clear that this procedure is far more flexible than the convex hull, although it retains the simplicity of the latter.
\vspace{2 mm}

\noindent
\emph{Characteristic hull polygon (CHP)} \vspace{2 mm}

% La informaci\'{o}n y el dibujo del characteristic hull los cambi\'{e} el 7-junio-2017. Hay que incorporarla al fichero definitivo.

As the LoCoH, the characteristic hull polygon (Downs and Horner 2009) is a union of convex sets and a generalization of the MCP. First, the Delaunay triangulation of the sample of locations is constructed. Then, the largest 5\% of the triangles are removed, where the size of the triangle is measured by its perimeter. As Downs and Horner (2009) point out, it would be interesting that the proportion of triangles to remove should somehow depend on the actual shape of the real home range (for instance, this proportion should be 0 for convex domains).

\begin{figure}[h]
\begin{center}
\includegraphics[width=9cm]{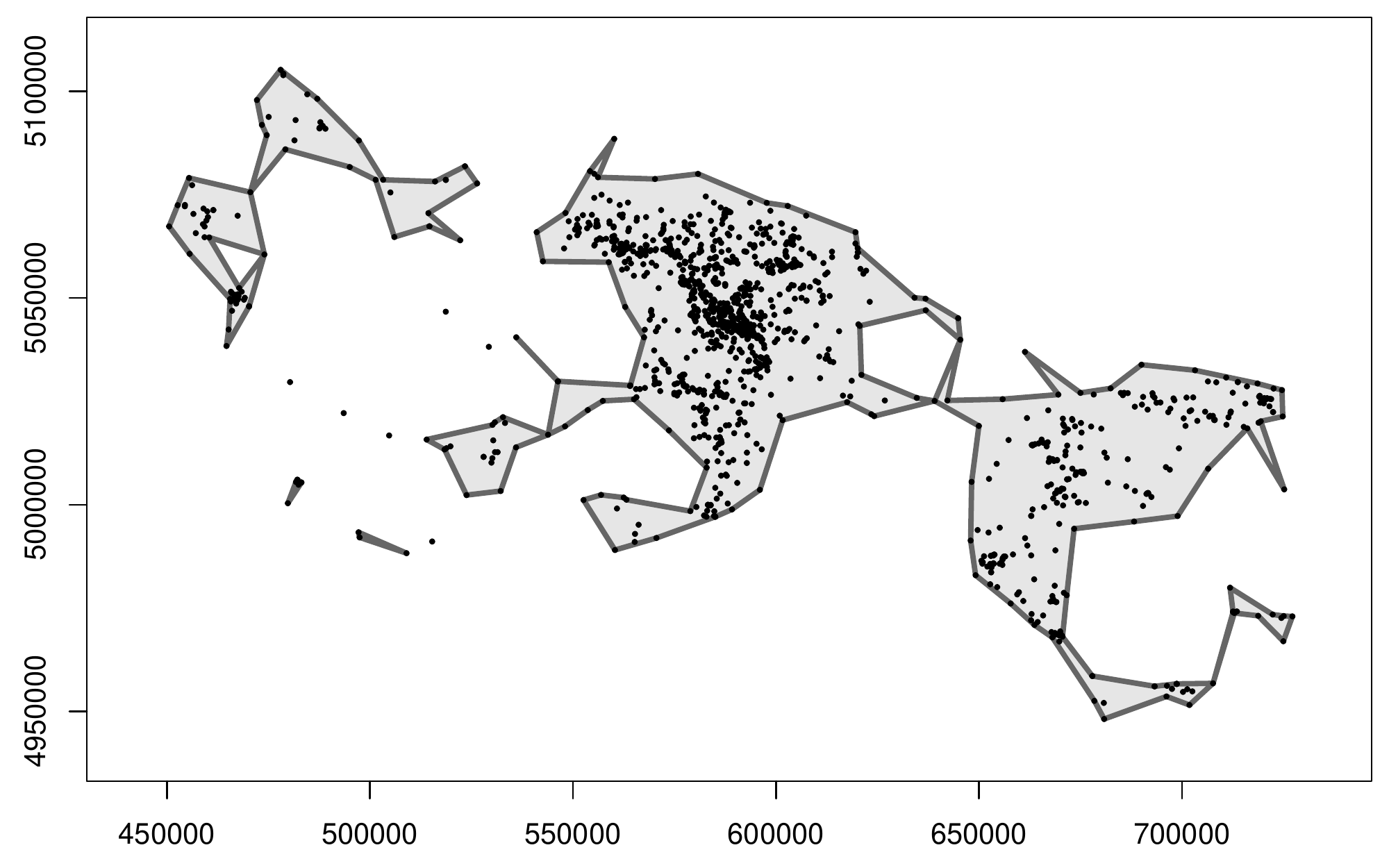}
\end{center}
\caption{Characteristic hull polygon for Zimzik locations.}
\label{Figure.Zimzik.CHP}
\end{figure}

The package {\tt adehabitatHR} is the only one implementing the CHP in R, but it measures the size of the triangles by their area. We have modified the {\tt CharHull} function from {\tt adehabitatHR} to compute the perimeter of the Delaunay triangles and remove the 5\% with largest boundary (see Figure~\ref{Figure.Zimzik.CHP}). We see that this estimator resembles the LoCoH with $k=35$. As a home range, the CHP is unsatisfactory, since it includes several slivers corresponding to seldom visited locations, and, in particular, this produces two surprising holes that do not seem reasonable for the home range (by visual inspection of the sample shape and taking into account the geography of the terrain). \vspace{2 mm}

\noindent
\emph{Single-linkage cluster} \vspace{2 mm}

Kenward {\em et al.} (2001) proposed yet another localized variant of the MCP. The difference with the LoCoH lies in the way that local neighbourhoods are constructed. Their proposal starts from the nearest-neighbour distances between locations, these distances are clustered using single-linkage cluster analysis, aiming to minimize the mean joining distance and imposing a certain minimum cluster size. Once the locations are grouped in clusters, the convex hull (or the minimum area polygon) of each cluster is computed and the final home range is defined as the union of these polygons.

\begin{figure}[h]
\begin{center}
\includegraphics[width=9cm]{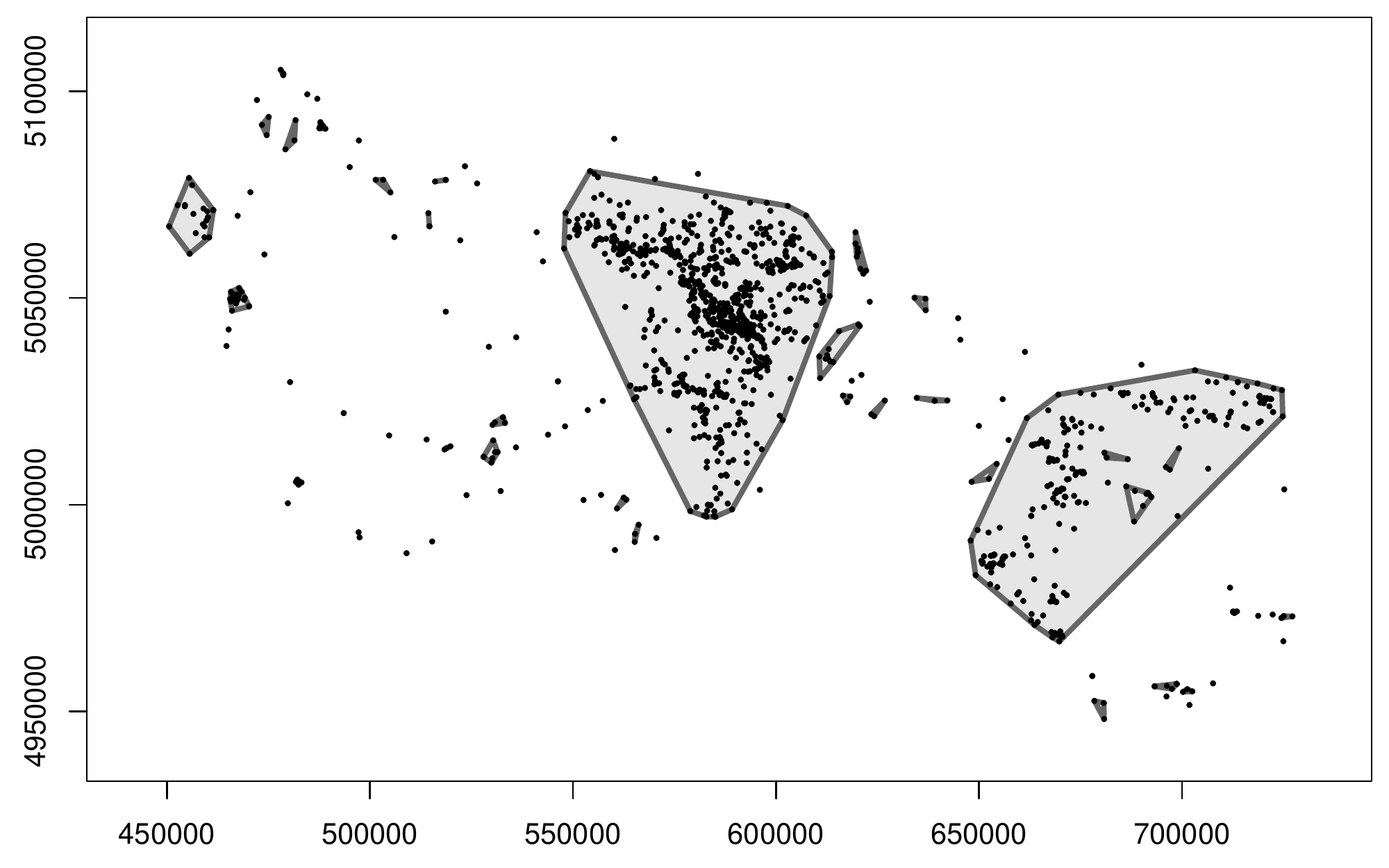}
\end{center}
\caption{Single-linkage cluster home range estimation for Zimzik locations.}
\label{Figure.Wolves.SLC}
\end{figure}

The single-linkage cluster home range of Zimzik locations data appears in Figure~\ref{Figure.Wolves.SLC}, produced using the package {\tt adehabitatHR}. The method identified two larger clusters of locations, and several other smaller clusters, and obtained the final home range estimate by gathering together the convex hulls of each of the clusters. Nevertheless, it is noticeable that the convex hull does not seem to appropriately recover the shape of the main clusters, since the latter do not appear to be convex.
\vspace{2 mm}

\subsection{Incorporating time dependency} \label{Subsection.TimeDependentHR}

As noted in Subsection~\ref{Subsection.AnimalLocations}, the advent of GPS technologies to the animal tracking context has posed new interesting problems. The high frequency in the signal transmission originates stochastic dependence among locations which are close in time. This dependence should somehow be incorporated into the home range estimator, but it is also a source of information on the behavioural mechanisms of the animal.

In this subsection we review the proposals to incorporate the time dependency into the home range estimator. The sample including the observation times is denoted by $(t_1,\bx_1),\ldots,(t_n,\bx_n)$, with $t_1<\ldots<t_n$. {It should be stressed that for most studies the observation times are fixed by the researcher, so that they should not be treated as random.}

We should also remind that static home range estimators are usually defined as a high probability contour of the utilization distribution. Most dynamic home range estimation methodologies seem to be also oriented towards this natural population goal, but some caution is required regarding recently proposed dynamic techniques, since some of them seem to lose track of the target that they intend to estimate.
%\vspace{2 mm}

\subsubsection{Global methods} \label{Subsubsection.GlobalMethodsWithTime}

\noindent
\emph{Time Geography Density Estimation (TGDE)} \vspace{2 mm}

Time geography is a perspective of social geography introduced by H\"{a}gerstrand (1970). It analyzes human phenomena (transportation, socio-economic systems,\ldots) in {a} geographic space taking into account the restrictions and trade-offs in human spatial behaviour imposed by individual activity schedules (see Miller 2005).

\begin{figure}[h]
\begin{center}
\includegraphics[trim={4.7cm 8.2cm 4.9cm 4.5cm},clip,width=7cm]{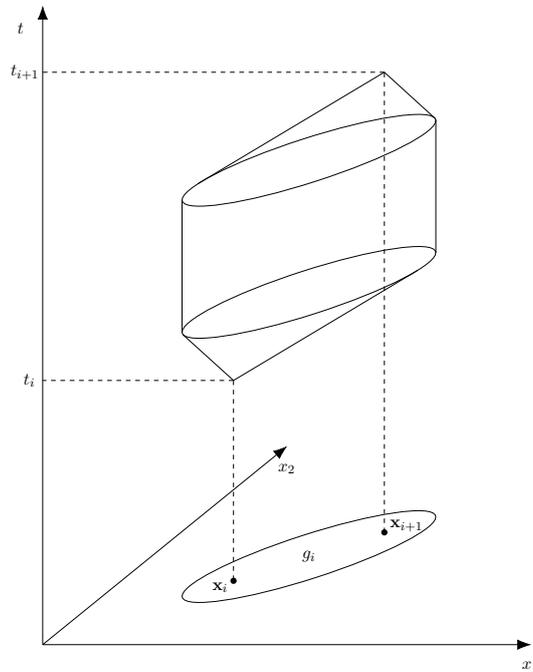}
\end{center}
\caption{Space-time prism and geoellipse $g_i$ corresponding to consecutive locations $\bx_i$ and $\bx_{i+1}$, observed at times $t_i$ and $t_{i+1}$ respectively.}
\label{Figure.SpaceTimePrism}
\end{figure}

An important concept in time geography is the space-time prism, which bounds all possible space-time paths between two observed locations $\bx_i$ and $\bx_{i+1}$ at consecutive time points, taking into account the accessibility constraints of the individual (see Figure~\ref{Figure.SpaceTimePrism}). The {\em geo-ellipse} $g_i$ or {\em potential path area} (PPA) of these two consecutive locations delineates in {the} geographic space all points that the individual can potentially reach during the time interval $[t_i,t_{i+1}]$:
$$
g_i = \{\bx: \|\bx-\bx_i\| + \|\bx-\bx_{i+1}\| \leq (t_{i+1}-t_i) v_i \},
$$
where $v_i$ denotes the maximum velocity in the time interval.

Downs (2010) and Downs {\em et al.} (2011) propose to integrate KDE and time geography techniques by using the geo-ellipses derived from location pairs as surrogates for the KDE level sets. Downs (2010) claims that a drawback of KDE home range estimation is that it includes areas where the individual could not have been located, given the spatial and temporal constraints established by the observed locations and the maximum potential velocity. As an alternative, if the maximum interval velocities $v_i$ are all assumed to be equal to some global maximum velocity $v$, then the time-geographic density estimate at a point $\bx$ is defined as
\begin{equation} \label{TGDE}
\hat f_{\mbox{\tiny TG}}(\bx) = \frac{1}{(n-1)[(t_n-t_1) v]^2} \sum_{i=1}^{n-1} G\left( \frac{\|\bx-\bx_i\|+\|\bx-\bx_{i+1}\|}{(t_{i+1}-t_i) v} \right),
\end{equation}
where $G$ is a decreasing function playing the role of the kernel. Given that $G$ is maximal at zero, the individual summands of the estimator (\ref{TGDE}) assign the highest probability to the points along the straight path between two consecutively observed locations and spread out the remaining probability mass in ellipsoidal contours having the observed locations as their foci. The velocity $v$  plays the role of the smoothing parameter and, as such, it is reasonable to derive its value from the animal locations and their corresponding observation times (see Long and Nelson 2012). The TGDE home range is the 95\% level set, $\{\hat f_{\mbox{\tiny TG}} \geq \hat c_{0.05}\}$, of the density estimate given in (\ref{TGDE}), where
$$
0.95 = \int_{\{\hat f_{\mbox{\tiny TG}}\geq \hat c_{0.05}\}} \hat f_{\mbox{\tiny TG}}(\bx)\, d\bx.
$$
%{Y no ser\'\i a m\'{a}s sencillo tomar el HR estimado con un KDE y, simplemente, intersecarlo con el PPA para {as\'\i}  excluir las zonas donde el animal, por restricciones de velocidad, no puede haber llegado?}

In parallel, Long and Nelson (2012) defined the PPA home range as the union of the $n-1$ geo-ellipses obtained from all the pairs of consecutive locations in the sample:
\begin{equation} \label{PPAHR}
\PPA_{\mbox{\tiny HR}} = \bigcup_{i=1}^{n-1} g_i,
\end{equation}
with $v_i=v$, for all $i=1,\ldots,n-1$. The estimator (\ref{PPAHR}) is a particular case of the TGDE home range $\{\hat f_{\mbox{\tiny TG}} \geq \hat c_{0.05}\}$ with a uniform kernel $G$.
%Specifically, once the segment velocities $v_{i,i+1}=(x_{i+1}-x_i)/(t_{i+1}-t_i)$, $i=1,\ldots,n-1$, are computed, Long and Nelson (2012) suggest several estimates $\hat v_{\max}$ based on estimates of the upper bound of a random variable.

Long and Nelson (2015) introduce the dynamic PPA home range (dynPPA), an adaptation of the PPA estimator to possibly different motion states. It is obtained by allowing the mobility parameter $v$ to change over time. In practice, these authors divide the trajectory of the animal in dynamic phases and obtain an estimator of $v$ in each of them.

\begin{figure}[h]
\begin{center}
\begin{tabular}{cc}
\includegraphics[width=7.5cm]{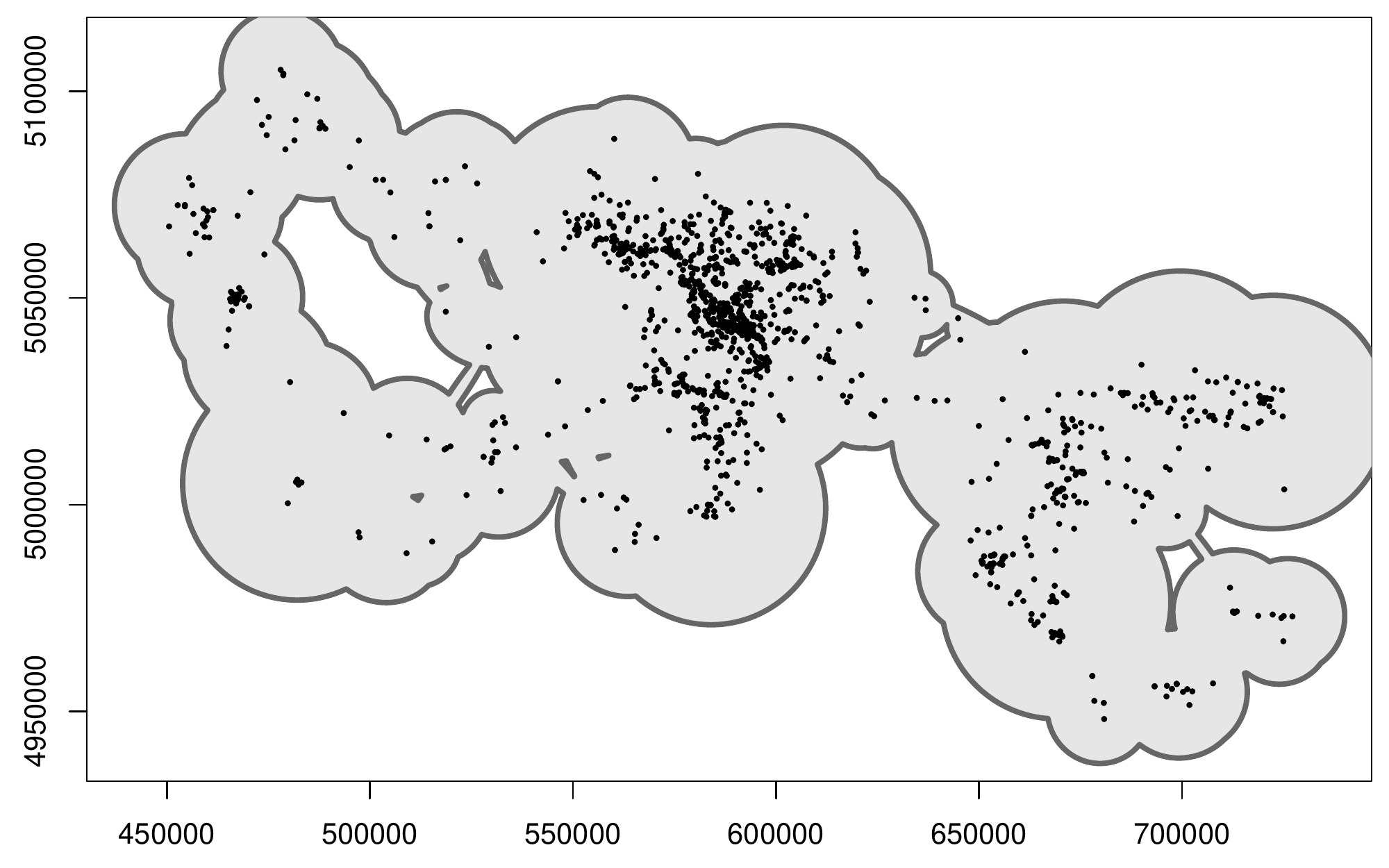} & \includegraphics[width=7.5cm]{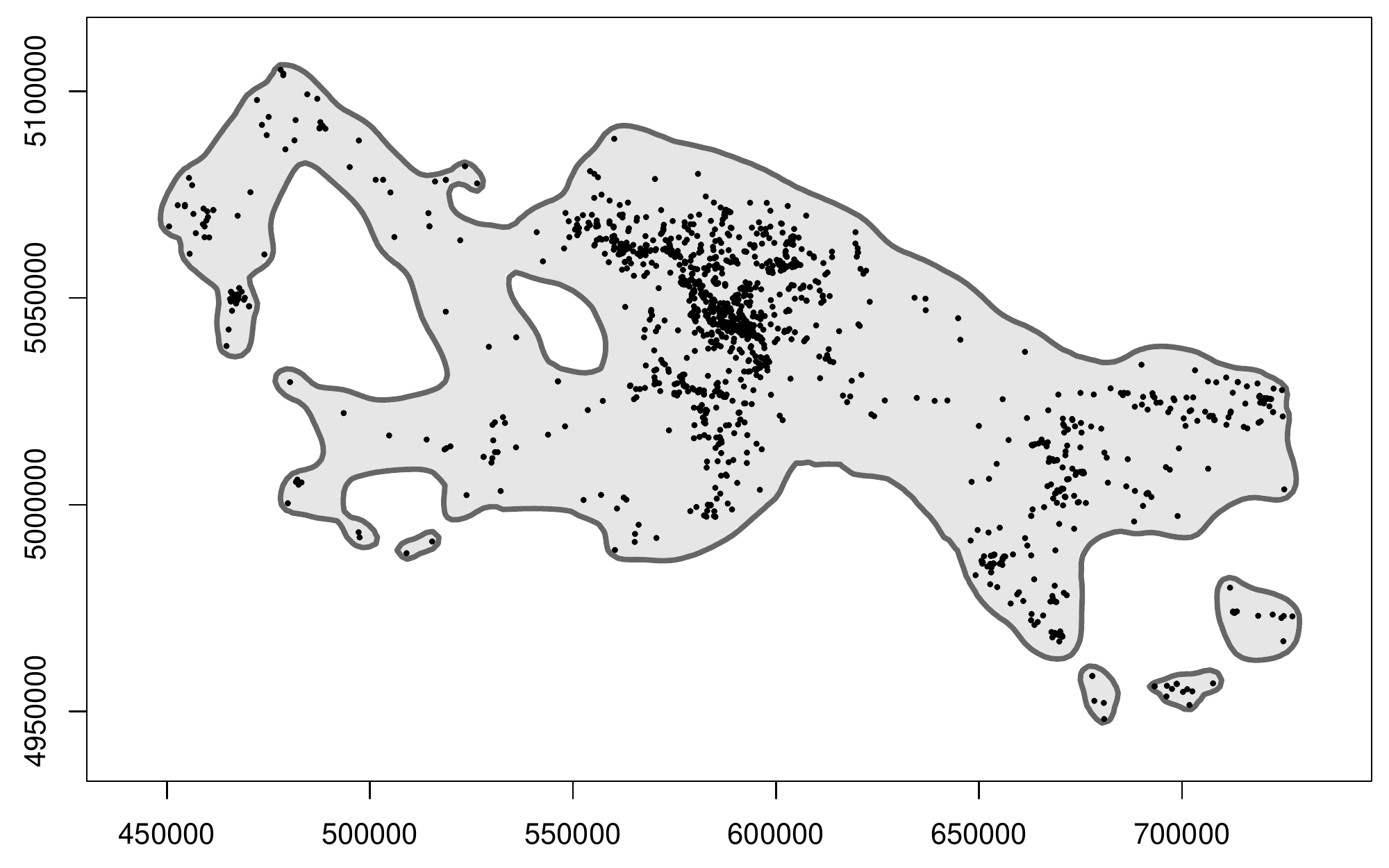} \\ (a) & (b)
\end{tabular}
\end{center}
\caption{Home ranges for Zimzik data using (a) PPA and (b) TGDE with Gaussian kernel.}
\label{Figure.TG}
\end{figure}

The PPA home range and the TGDE are implemented in the R package {\tt wildlifeTG} (Long 2017). In Figure~\ref{Figure.TG} we show both home range estimators for the Zimzik locations data, where the function $G$ is Gaussian-like (i.e., proportional to $e^{-x^2/2}$) for the TGDE in (b). The PPA home range represents, in a sense, an upper bound of the home range since it clearly overestimates it by including all the locations that the animal could have reached if it had moved from each of the observed locations at a certain speed. TGDE is clearly the most satisfactory of the two, although it comprises many seldom visited areas.

\vspace{2 mm}

\noindent
\emph{Kernel density estimation} \vspace{2 mm}

There have been various attempts to generalize the kernel home range estimator to incorporate the time dependence between the observed locations. Nevertheless, there are two important issues that should be remarked: first, the definition of the kernel density estimator for dependent data is exactly the same as for independent data, and second, regarding the fundamental problem of bandwidth selection, the data can be treated as if they were independent, since the asymptotically optimal bandwidth for independent data is also optimal under quite general conditions of dependence, as shown in Hall {\em et al.} (1995). This means that, to design methods to estimate the utilization distribution density, we can proceed exactly the same as for independent data.

Keating and Cherry (2009) suggested a product kernel density estimator where time was incorporated as an extra variable to the two-dimensional location vector, thus yielding three-dimensional observations. {This approach does not seem appropriate,} since time is not a random variable whose frequency we want to analyze{, as we noted at the beginning of Section 2.2}.

\begin{figure}[h]
\begin{center}
\includegraphics[width=9cm]{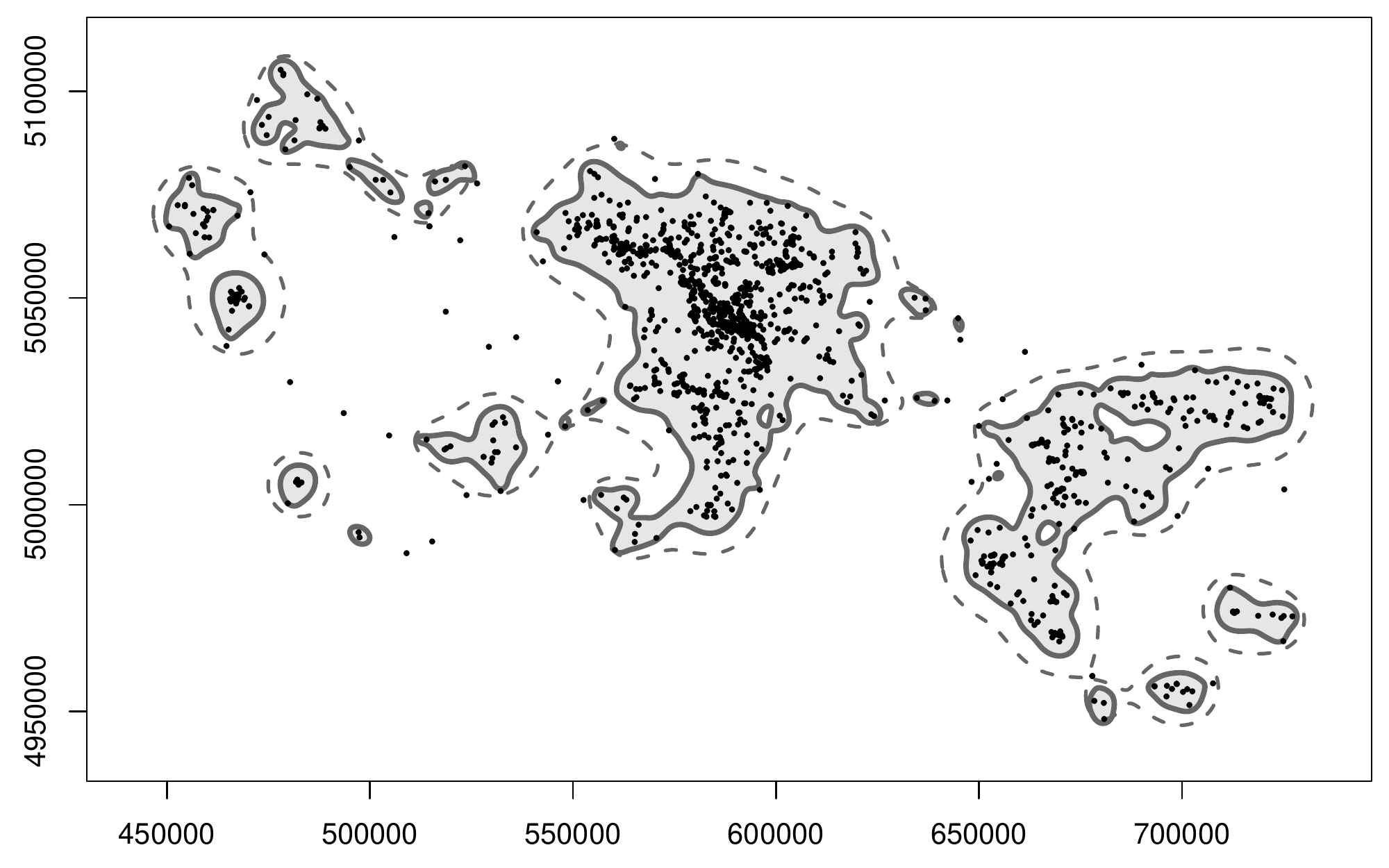}
\end{center}
\caption{MKDE home range with $h_{\min}=1$ (continuous line) and $h_{\min}=5000$ (discontinuous).}
\label{Figure.WolvesMKDE}
\end{figure}

In the context of estimating the {\em active} utilization distribution (describing space frequency use in the active moments of the animal), Benhamou and Cornelis (2010) developed the movement-based kernel density estimation (MKDE) method. MKDE consists in dividing each step or time interval $[t_i,t_{i+1}]$ into several sub-steps, that is, adding new points at regular intervals on each step. Then KDE is carried out on the known and the interpolated relocations with a variable one-dimensional smoothing parameter $h_i(t)$. For each time interval $h_i$ is a smooth function of the time lapse from $t_i$ and to $t_{i+1}$, taking its smallest value $h_{\min}$ at the end points and the largest (at most $h_{\max}$) at the midpoint. A drawback of MKDE is that it depends, thus, on the choice of several parameters, such as $h_{\min}$, $h_{\max}$ and the length of the subintervals. For instance, using package {\tt adehabitatHR}, in Figure~\ref{Figure.WolvesMKDE} we have plotted the MKDE home ranges for two very different values of $h_{\min}$ but equal values of the rest of parameters: clearly, the choice of this smoothing parameter can substantially alter the resulting home range. Optimal simultaneous selection of all the parameters of MKDE with respect to some criterion is computationally unfeasible even for moderate sample sizes.
A second concern regarding MKDE is that it is not clear whether the resulting home range differs substantially from the KDE based on the independence assumption.

There have been extensions to the original MKDE proposal. For instance, to incorporate knowledge of boundaries that the animal does not go through (rivers, cliffs, \ldots), Benhamou and Cornelis (2010) suggest to reset to 0 beyond the boundary and reflect with respect to the boundary the resulting estimate of the utilization density. Also, Tracey {\em et al.} (2014) use the MKDE on 3-dimensional location data.

Steiniger and Hunter (2013) propose the {\em scaled line kernel home range} estimator (SLKDE), which is similar to the MKDE of Benhamou and Cornelis (2010). Each line segment connecting two consecutive relocations, $\bx_i$ and $\bx_{i+1}$, is divided into subsegments and a KDE is constructed on each line segment based on observed and interpolated relocations. This estimator is called the {\em raster} $r_{i,i+1}$. Then all the rasters are pooled to construct an unnormalized estimator of the UD (it does not integrate to 1). The main difference with MKDE is that the smoothing parameter of the SLKDE is selected in such a way that it ``inflates'' the home range estimator on the observed relocations and thins it in the lines connecting them: the rasters have what Steiniger and Hunter (2013) call a ``bone-like'' shape. This procedure is not implemented in R. \vspace{2 mm}

%Based on their statement that the KDE home ranges resulting from the assumption of independence adjust too much to the shape of the trajectory,
Fleming {\em et al.} (2015) propose the {\em autocorrelated kernel density estimator} (AKDE). As usual, the probability density $f$ of the UD is estimated by a bivariate KDE $\hat f$ with a Gaussian kernel $K$ and an unconstrained smoothing parameter $\bH$. {As for the ad hoc bandwidth proposal of Worton (1989), the AKDE uses a bandwidth which is derived from the assumption of Gaussianity of the underlying distribution (which naturally leads to oversmoothing in most practical scenarios), with the only difference that instead of assuming that the location data are independent, they are supposed to be observations from a Gaussian stochastic process.}

%, chosen to minimize the mean integrated squared error (MISE) $\bbE\int(\hat f-f)^2$. The only difference with respect to the case of independent observations is that the unknown density $f$ in the expression of the MISE is claimed to be replaced by the density of a Gaussian stationary stochastic process.

\subsubsection{Localized methods} \label{Subsubsection.LocalizedMethodsWithTime}

\noindent
\emph{T-LoCoH} \vspace{2 mm}

T-LoCoH (Lyons {\em et al.} 2013) generalizes the LoCoH home range to incorporate the observation times of the relocations.
%``T-LoCoH integrates time with space in the construction of local hulls through a scaling that relates distance and time in reference to the individual's characteristic velocity''.
T-LoCoH incorporates the time associated to each location in two phases of the LoCoH algorithm: nearest neighbour selection and the ranking of hulls. First, the NN selection relies on the so-called time-scaled distance (TSD), which transforms the time coordinate into a third one of Euclidean space $\bbR^3$.
Specifically, the TSD between two sample points, {$(t_i,\bx_i)$ and $(t_j,\bx_j)$,} is defined as
$$
\TSD_{ij}^2 = \|\bx_i-\bx_j\|^2 + s^2 v_{\max}^2 (t_i-t_j)^2,
$$
where $s\geq 0$ is a scaling factor of the maximum theoretical velocity $v_{\max}$.
Finally, to construct the isopleths, local hulls are sorted according to a hull metric, chosen to reflect the spatial or time information we might want to use. Dougherty {\em et al.} (2017) suggest a cross-validation procedure to select the tuning parameters $k$ and $s$ for the T-LoCoH.

\begin{figure}[h]
\begin{center}
\includegraphics[width=9cm]{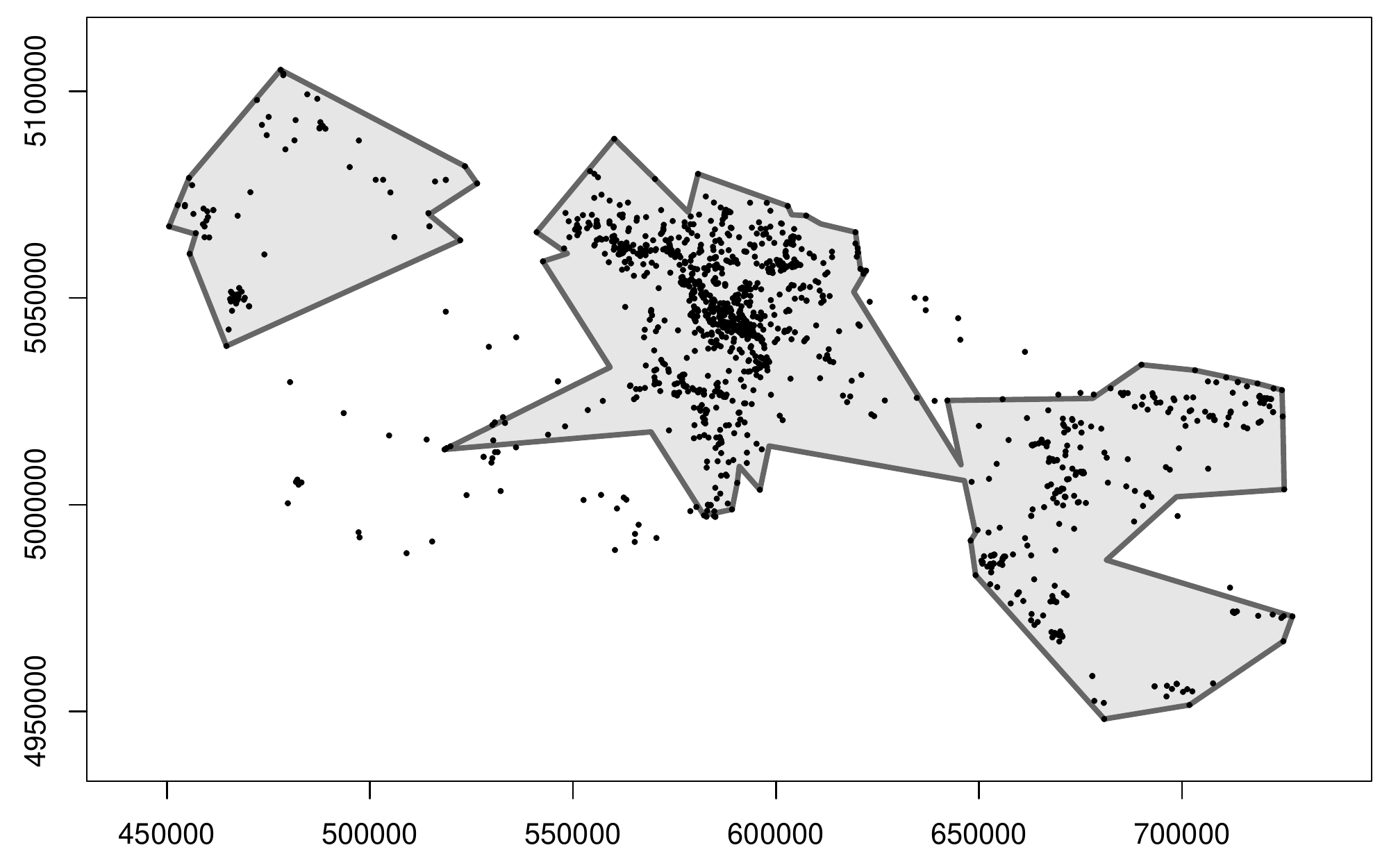}
\end{center}
\caption{T-LoCoH isopleths for probability content 95\%.}
\label{Figure.Wolves.TLoCoH}
\end{figure}

Following the indications in Lyons {\em et al.} (2013), with the aid of the graphical procedures available in the package {\tt tlocoh}, we have chosen the values of $s=0.05$ and $k=40$ as T-LoCoH parameters for Zimzik locations. The value of $v_{\max}$ was internally chosen by R.
Figure~\ref{Figure.Wolves.TLoCoH} displays the resulting T-LoCoH home range. We notice that, in this case, there is not such a big difference between the LoCoH homerange (Figure~\ref{Figure.Wolves.LoCoH}) and the T-LoCoH one, probably due to the small value of the parameter $s$.
\vspace{2 mm}

%AKDE is implemented in the R package {\tt ctmm}, but I am not able to make it work.

%Hall et al. (1995), Walter et al. (2015) state that data dependence does not influence much on kde. See Barbeito and Cao (2016). But they assume stationarity of the data.

\section{Ranking the home ranges} \label{Section.Selection}

One of the key questions after computing several home range estimates based on the same positional data is how to select the most convincing or realistic one. In most case studies the biological expert chooses the soundest result according to previous information on the animal or its species. The possibility of making an automatised choice, based on objective statistical criteria, among a collection of home range estimates  is, in a sense, still an open question.

There have been some attempts to compare home ranges according to different criteria. Here we briefly review the proposals valid for real data (methods only working for simulated samples are not considered). One possibility, especially if locations are treated as independent data, is to separate the sample into two subsamples of locations: a training sample to construct the home range estimator and a test sample to check its predictive accuracy. Approaches based on this idea have been {used, for instance, in Kranstauber {\em et al.} (2012) and in Tarjan and Tinker (2016).} Fleming {\em et al.} (2015) compare the areas of the home range estimates obtained with the complete sample and with the first half of the data (chronologically speaking). {Long and Nelson (2015) compute the areas of different home range estimates. Kenward {\em et al.} (2001) study the relationship between the logarithm of the home range area and, for instance, environmental factors known to influence the animal behaviour (such as food availability or population density).
Apart from the area, Steiniger and Hunter (2013) use shape complexity as given, for example, by holes and patches and the presence or not of corridors in the home range.}

Cumming and Corn\'{e}lis (2012) and Walter {\em et al.} (2015) compare home range estimates using the area-under-the-curve (AUC) corresponding to {receiver operating characteristic (ROC)} curves. The ROC curve (one for each HR) crucially depends on the choice of a certain raster containing all the home ranges. The curve is computed on the basis of the probability assigned by the UD to each cell in the raster and of the label indicating whether there is any sighting in the cell or not. %Observe that this choice of the labels may favour larger home ranges.
% No se entiende nada de las explicaciones que dan estos autores respecto a c\'{o}mo calculan la ROC curve.

In this work we propose a new automatic way of choosing the ``best fitting'' home range among a collection, or at least a criterion for ranking them. The idea is to minimize a measure of the excess extension incurred by the home range as compared with the original locations, penalized by a measure of sample overfitting. The procedure is valid for any type of home range, {regardless of the way it was constructed (for instance, estimating a utilization density or through geometric set estimation procedures).}

\begin{figure}[h]
\begin{center}
\includegraphics[width=12cm]{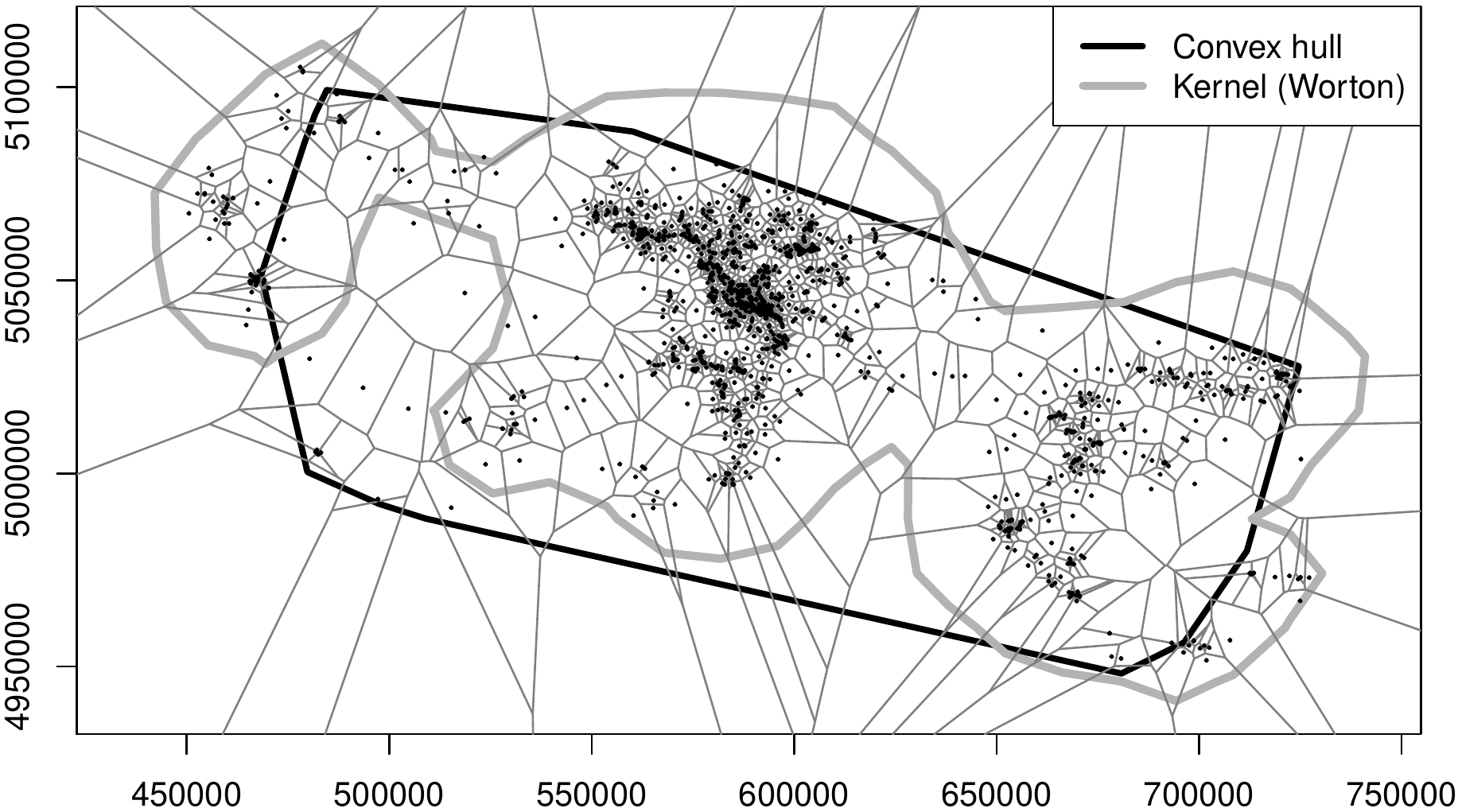}
\end{center}
\caption{Voronoi tessellation of Zimzik relocations and two superimposed home ranges.}
\label{Figure.SuperimposedVoronoi}
\end{figure}

\begin{figure}[h]
\begin{center}
\begin{tabular}{@{}c@{}c@{}}
\includegraphics[width=8cm]{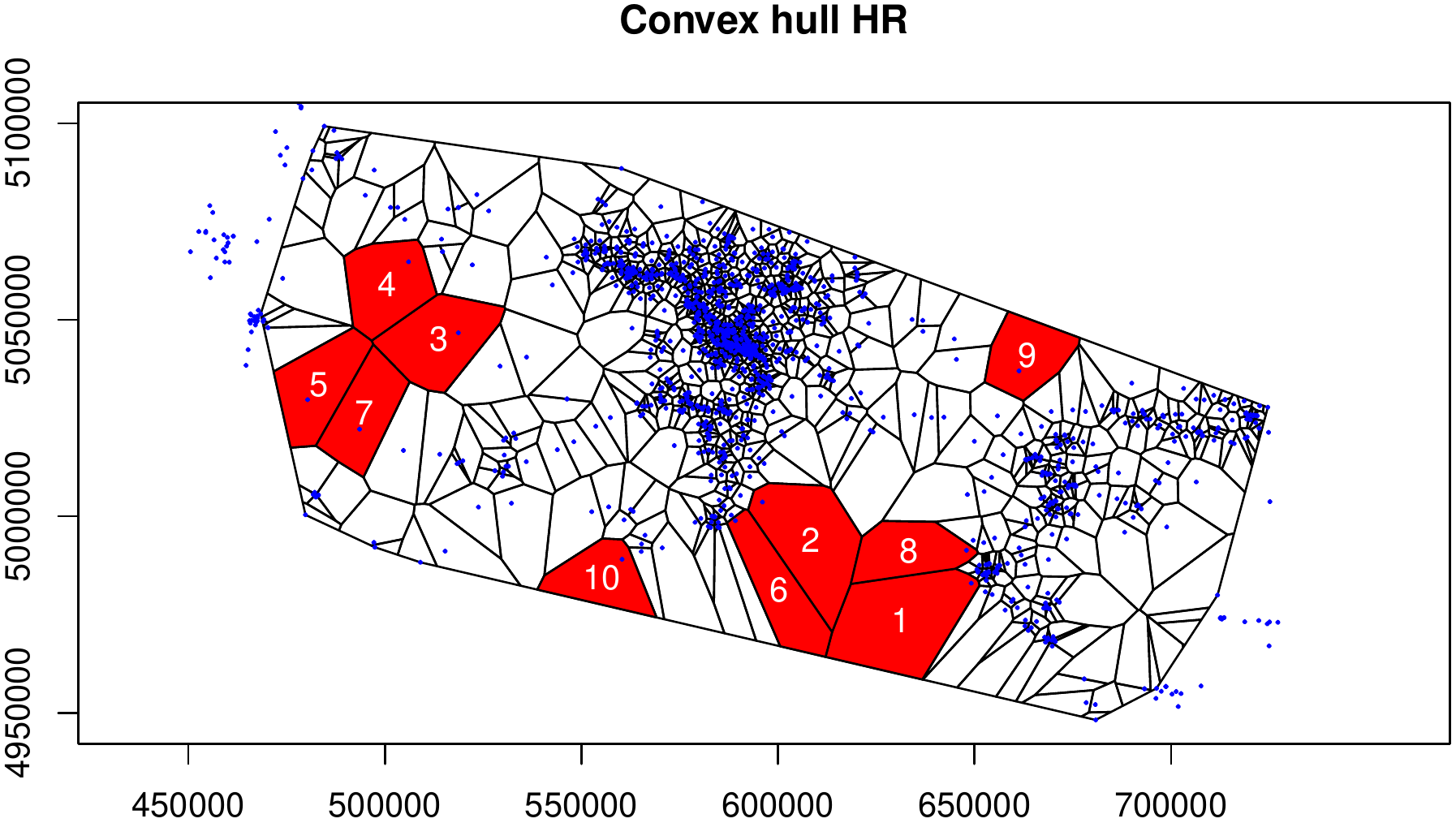} & \includegraphics[width=8cm]{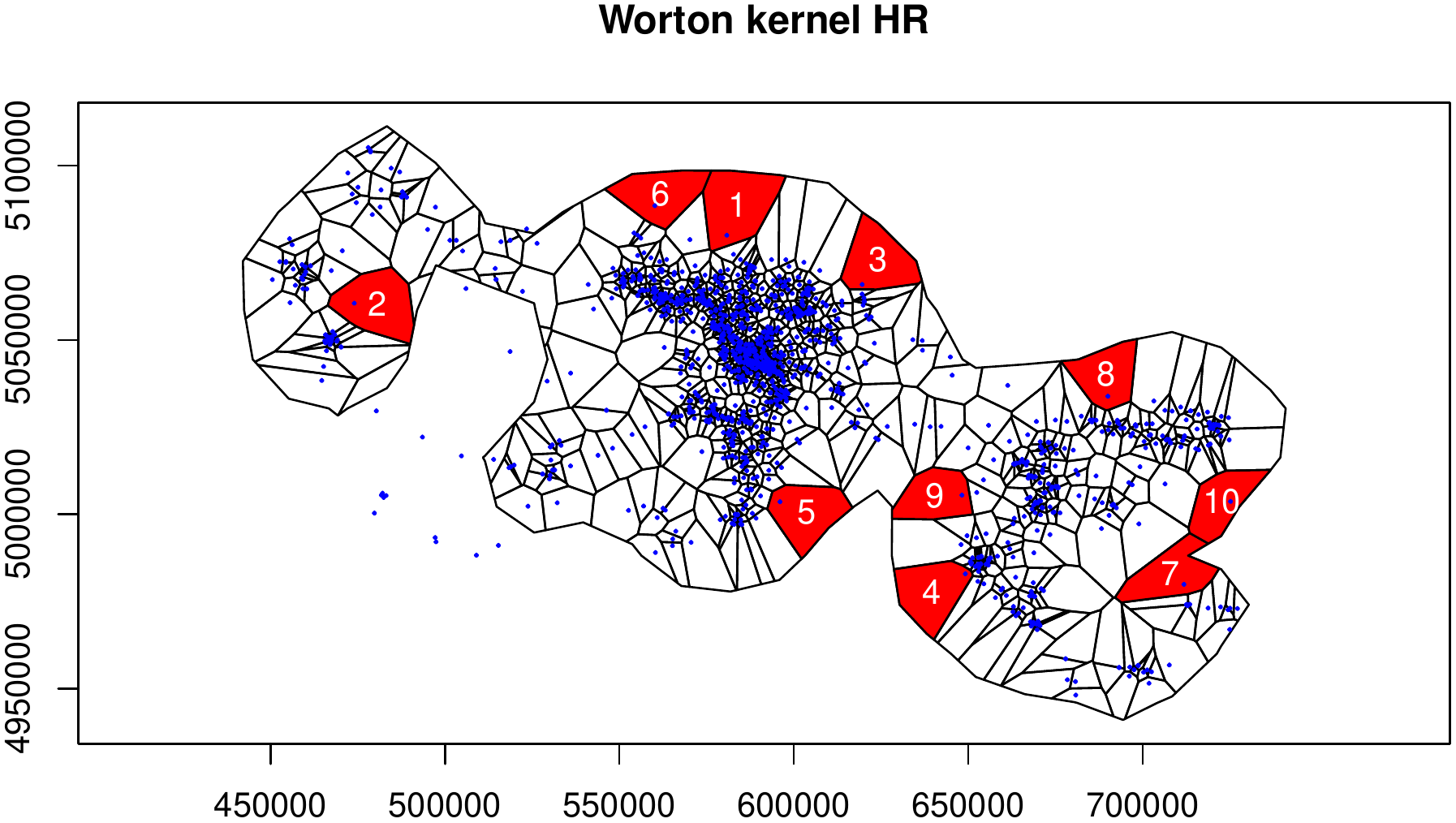} \\
\includegraphics[width=8cm]{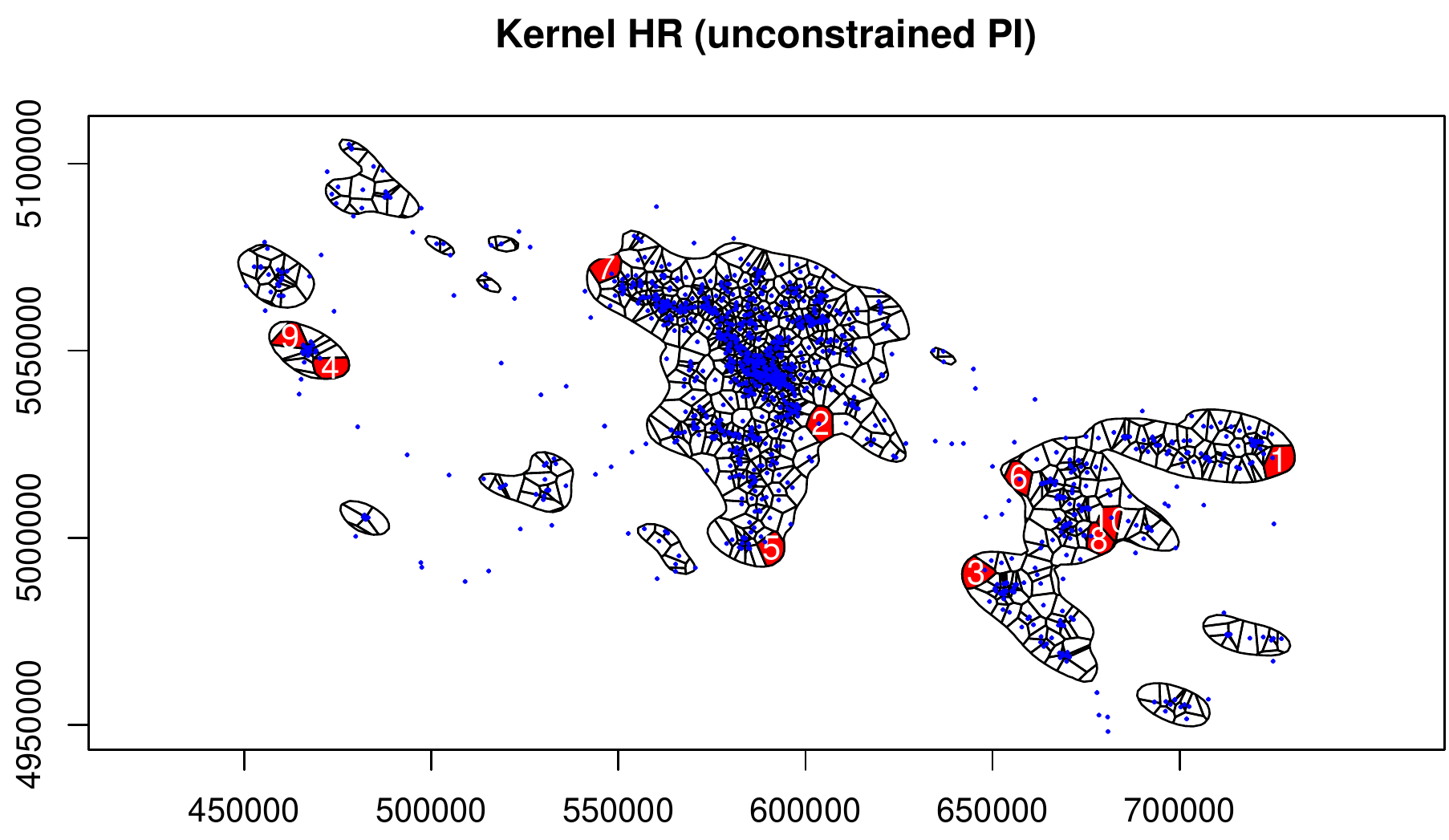} & \includegraphics[width=8cm]{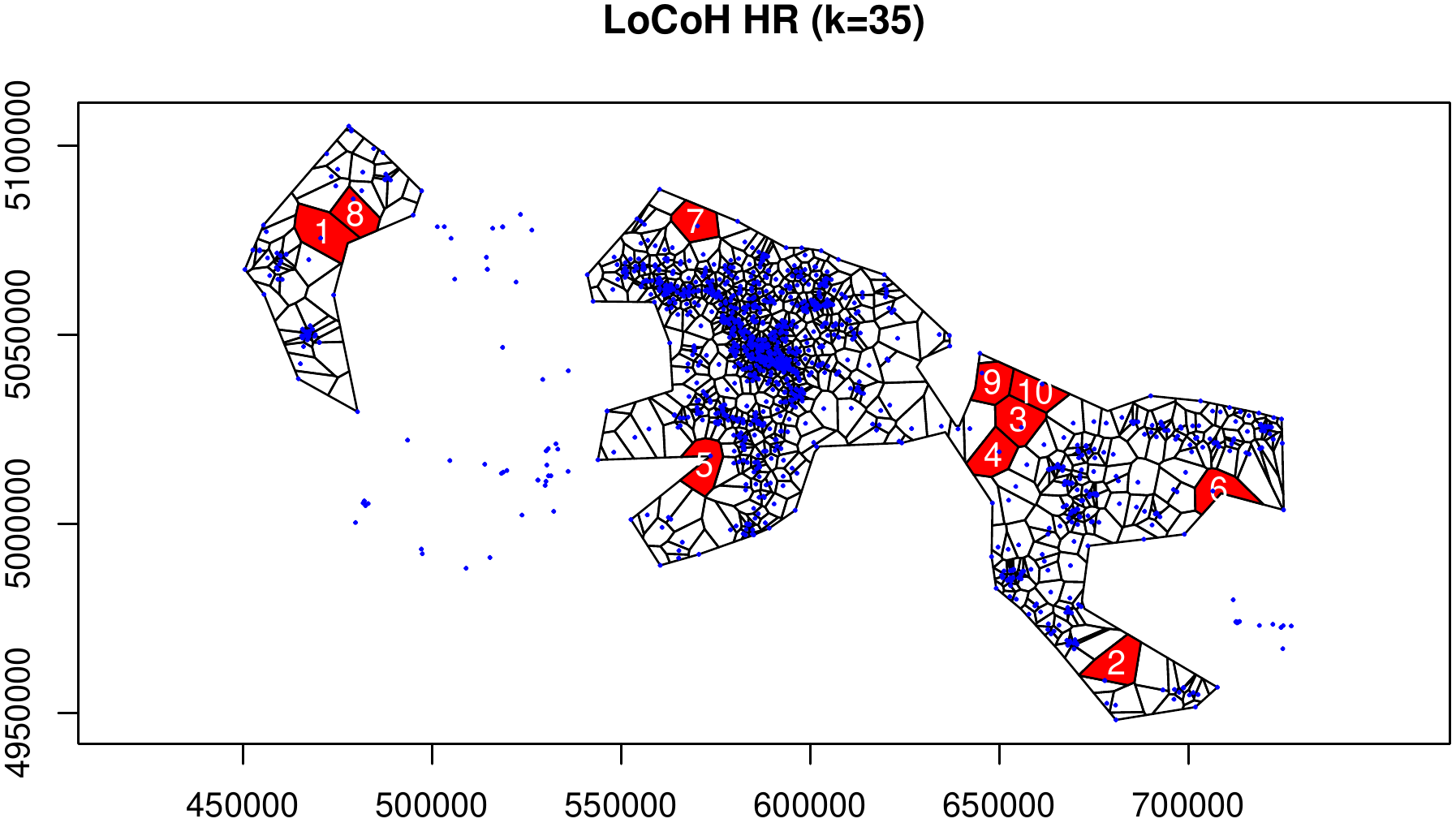}
\end{tabular}
\end{center}
\caption{Intersection of Voronoi tessellation with some home ranges and highlight of the ten largest (intersected) cells sorted and numbered by area.}
\label{Figure.LargestVoronoiCells}
\end{figure}

To quantify the excess area of a certain home range with respect to the observed sample,
we propose to intersect the Voronoi tessellation of the sample with that home range (see Figure~\ref{Figure.SuperimposedVoronoi}). Then we sort the resulting (intersected) Voronoi cells according to their area (see Figure~\ref{Figure.LargestVoronoiCells}). We denote by $S_{(i)}$ the area of the $i$-th largest cell after intersection with the home range. Observe that zones contained in the set estimator but never or seldom visited by the animal usually correspond to large (intersected) Voronoi cells. We take the area of the largest cells as a proxy for the measure of the home range ``hollowness''. Specifically we have computed $S_{(1)}$, the area of the largest Voronoi cell, and $\sum_{i=1}^{10} S_{(i)}$, the area of the ten largest cells. Other approaches such as taking a fixed proportion of the cells (i.e., of the sample size) could be considered.

% P\'{a}rrafo reescrito el 9-jun-2017. Hay que incorporarlo en nuestro documento final.
The home range area $A$, the maximum $S_{(1)}$ and the sum of the largest ten $\sum_{i=1}^{10} S_{(i)}$ areas of the Voronoi cells are displayed in Table~\ref{Table.ErrorMeasures} for a selection of home ranges based on Zimzik data. All areas are in $m^2$. The home ranges were selected for illustration purposes by their simplicity and/or reasonable graphical similitude to the original tracking data. The considered estimators are: the convex hull (containing only 95\% of the sample points), the LoCoH with $k=35$ and $k=80$, the kernel estimator with the ``ad hoc'' smoothing parameter $h = 12269.09$ used by Worton (1989) and with the unconstrained plug-in bandwidth matrix (\ref{UnconstrainedPIbandwidth}), the T-LoCoH with $k=40$ and $s=0.06$ and the MKDE with $h_{\min}=12000$ (a value slightly smaller than Worton's bandwidth). We have also computed the area $B=\mbox{48,402,728,426}$ of the smallest bounding box containing all these home ranges. In Table~\ref{Table.RatiosAreas} we display the ratios of $S_{(1)}$ and $\sum_{i=1}^{10} S_{(i)}$ over $B$ (in \%), as a measure of how much the home range overestimates its target. As expected, {for all these four measures} the convex hull is the home range exhibiting the largest overestimation ratio, followed by the group formed by the LoCoH with $k=80$, the T-LoCoH, the kernel estimator with the ``ad hoc'' bandwidth $h$ and the MKDE. The {kernel estimator with the plug-in bandwidth attains always the lowest overestimation ratio, followed by the LoCoH with $k=35$.}

% Tabla rehecha el 9-jun-2017, habiendo corregido los errores observados en el estimador kernel de adehabitatHR y los errores en la selecci\'{o}n del h plug-in debido a las unidades de los datos.
% Hay que incorporarla en nuestro documento final.
\begin{table}[h]
\begin{center}
\begin{tabular}{@{}l|r|rr|}
 \multicolumn{1}{c}{} & \multicolumn{1}{c}{} & \multicolumn{2}{c}{Area of Voronoi cells} \\ \cline{3-4}
 \multicolumn{1}{c}{Home range} & \multicolumn{1}{c}{Total area $A$} & \multicolumn{1}{c}{$S_{(1)}$} & \multicolumn{1}{c}{$\sum_{i=1}^{10} S_{(i)}$} \\ \hline
Convex hull & 24,419,954,019 & 719,264,280 & 4,445,094,315 \\
LoCoH ($k=35$) & 10,851,131,941 & 148,388,790 & 1,078,176,739 \\
LoCoH ($k=80$) & 18,837,169,195 & 338,525,431 & 2,509,247,802 \\
Kernel (Worton) & 15,982,259,965 & 244,709,734 &  1,787,449,239 \\
Kernel (plug-in) & 8,083,494,038 & 53,206,494 & 455,072,754 \\ % Es raro: el estimador sale muy peque\~{n}o y fragmentado
T-LoCoH ($k=40$, $s=0.06$) & 12,768,423,611 & 352,939,461 & 1,924,856,639 \\
MKDE & 22,944,329,048 & 330,576,812 & 2,638,851,646 \\\hline
\end{tabular}
\end{center}
\caption{Home range area $A$, maximum area $S_{(1)}$ and area of largest ten $\sum_{i=1}^{10} S_{(i)}$ intersected Voronoi cells for some home ranges.}
\label{Table.ErrorMeasures}
\end{table}

% Tabla rehecha el 9-jun-2017, habiendo corregido los errores observados en el estimador kernel de adehabitatHR y los errores en la selecci\'{o}n del h plug-in debido a las unidades de los datos. Adem\'{a}s hemos quitado las dos columnas en las que se divid\'{\i}a por A = \'{a}rea del HR.
% Hay que incorporarla en nuestro documento final.
\begin{table}[h]
\begin{center}
\begin{tabular}{@{}l|cc|}
 \multicolumn{1}{c}{Home range} & \multicolumn{1}{c}{$S_{(1)}/B$} & \multicolumn{1}{c}{$\sum_{i=1}^{10} S_{(i)}/B$} \\ \hline
Convex hull & 1.49 & 9.18 \\
LoCoH ($k=35$) & 0.31 & 2.23 \\
LoCoH ($k=80$) & 0.70 & 5.18 \\
Kernel (Worton) & 0.51 & 3.69 \\
Kernel (plug-in) & 0.11 & 0.94 \\
T-LoCoH ($k=40$, $s=0.06$) & 0.73 & 3.98 \\
MKDE & 0.68 & 5.45 \\\hline
\end{tabular}
\end{center}
\caption{Home range ratios (in \%) of the areas of the intersected Voronoi cells over the area $B$ of an enclosing box.}
\label{Table.RatiosAreas}
\end{table}

% Tabla rehecha el 9-jun-2017, habiendo corregido los errores observados en el estimador kernel de adehabitatHR y los errores en la selecci\'{o}n del h plug-in debido a las unidades de los datos.
% Hay que incorporarla en nuestro documento final.
\begin{table}[!th]
\begin{center}
\begin{tabular}{lrr} \hline
HR estimator & \multicolumn{1}{c}{$C$} & \multicolumn{1}{c}{$1/C$} \\ \hline
Convex hull & 0.7090 & 1.4104 \\
LoCoH ($k=35$) & 0.1510 & 6.6215 \\
LoCoH ($k=80$) & 0.4053 & 2.4674 \\
Kernel (Worton) & 0.2548 &  3.9251 \\
Kernel (plug-in) & 0.0723 & 13.8232 \\
T-LoCoH ($k=40$) & 0.1607 & 6.2229 \\
MKDE & 0.3101 & 3.2245 \\ \hline
\end{tabular}
\end{center}
\caption{Circularities for some home range estimators based on Zimzik relocations.}
\label{Table.DefectError}
\end{table}

% P\'{a}rrafo reescrito el 9-jun-2017. Hay que incorporarlo en nuestro documento final.
As a measure of how much a home range estimator overfits the underlying sample, we use {\em shape descriptors} (see, e.g., Gonz\'{a}lez and Woods 2008, Ch. 11). Specifically, since undersmoothing produces deep intrusions into the home range shape and consequently a perimeter increase, {we have computed the shape {\em circularity} $C$, also called form factor,} a simple descriptor taking values in (0,1]:
$$
C = \frac{4\pi \, \mbox{area}}{\mbox{perimeter}^2} .
$$
%This simple measure is closely related to the {\em nuclear contour index}, $\mbox{NCI}= \mbox{perimeter}/\sqrt{\mbox{area}} $, another shape descriptor widely employed to describe the morphology of tumors {\bf (citation needed)}.
In particular, $C$ takes a value of 1 for a circular shape, but it is affected by the aspect ratio, so it can be close to 0 both for a flattened ellipse and a shape full of spikes. In Table~\ref{Table.DefectError} we display the circularity for the same home range estimators as in Table~\ref{Table.ErrorMeasures}. The results are not surprising: the three estimators whose shape is more adapted to the sample (namely, LoCoH with $k=35$, the kernel estimator with plug-in bandwidth and the T-LoCoH) are the ones with lowest circularity ($<0.17$).

% P\'{a}rrafo reescrito el 9-jun-2017. Hay que incorporarlo en nuestro documento final.
The overestimation ratios (in \%), generically denoted by $R$, based on the Voronoi tessellation and the circularity $C$ are combined in a penalized criterion $R_p := R+\lambda/C$, where $\lambda>0$ is a tuning parameter regulating the trade-off between the home range size and its goodness of fit to the locations. For Zimzik data, in Table~\ref{Table.Penalization} we reproduce the values of $R_p$ for all the home ranges and ratios $R=100\cdot S_{(1)}/B$ considered in the previous tables and for different choices of $\lambda$. Interestingly, we see that the LoCoH estimate with $k=80$ is the home range minimizing the penalized criterion in most occasions. In Table~\ref{Table.Penalization2} we display the values of $R_p$ for $R=100\cdot\sum_{i=1}^{10} S_{(i)}/B$. Here it is the kernel home range, with the ad hoc choice of bandwidth, the one yielding better results, although the LoCoH with $k=35$ and $k=80$ has a similar performance. As a conclusion, we could say that the LoCoH (with a reasonable choice of $k$) and the kernel estimator proposed by Worton (1989) seem to be a reasonable and safe option in all circumstances. The MKDE is the time-dependent HR working best among those considered. Note, however, that the time-dependent estimators still have room for improvement since a more adequate choice of the smoothing parameters could diminish the criterion $R_p$.

% Tabla rehecha el 9-jun-2017, habiendo corregido los errores observados en el estimador kernel de adehabitatHR y los errores en la selecci\'{o}n del h plug-in debido a las unidades de los datos. Adem\'{a}s hemos quitado la fila en la que se divid\'{\i}a por A = \'{a}rea del HR.
% Hay que incorporarla en nuestro documento final.
\begin{table}[h]
\begin{center}
\begin{tabular}{lccccccc}
 &    & \multicolumn{2}{c}{LoCoH} & \multicolumn{2}{c}{Kernel} & T-LoCoH & \\
 & CH & $k=35$ & $k=80$ & Worton & plug-in & $k=40$ & MKDE \\ \hline
$\lambda=0.01$ & 1.500 & 0.373 & 0.724 & 0.545 & {\bf 0.248} & 0.791 & 0.715 \\ \hline
$\lambda=0.05$ & 1.557 & {\bf 0.638} & 0.823 & 0.702 & 0.801 & 1.040 & 0.844 \\ \hline
$\lambda=0.1$ & 1.627 & 0.969 & 0.946 & {\bf 0.898} & 1.492 & 1.351 & 1.005 \\ \hline
$\lambda=0.2$ & 1.768 & 1.631 & {\bf 1.193} & 1.291 & 2.875 & 1.974 & 1.328 \\ \hline
$\lambda=0.3$ & 1.909 & 2.293 & {\bf 1.440} & 1.683 & 4.257 & 2.596 & 1.650 \\ \hline
$\lambda=0.4$ & 2.050 & 2.955 & {\bf 1.686} & 2.076 & 5.639 & 3.218 & 1.973 \\ \hline
$\lambda=0.5$ & 2.191 & 3.617 & {\bf 1.933} & 2.468 & 7.022 & 3.841 & 2.295 \\ \hline
\end{tabular}
\end{center}
\caption{$R_p$ values for $R=100\cdot S_{(1)}/B$.}
\label{Table.Penalization}
\end{table}

% Tabla rehecha el 9-jun-2017, habiendo corregido los errores observados en el estimador kernel de adehabitatHR y los errores en la selecci\'{o}n del h plug-in debido a las unidades de los datos. Adem\'{a}s hemos quitado la fila en la que se divid\'{\i}a por A = \'{a}rea del HR.
% Hay que incorporarla en nuestro documento final.
\begin{table}[h]
\begin{center}
\begin{tabular}{lccccccc}
 &    & \multicolumn{2}{c}{LoCoH} & \multicolumn{2}{c}{Kernel} & T-LoCoH & \\
 & CH & $k=35$ & $k=80$ & Worton & plug-in & $k=40$ & MKDE \\ \hline
$\lambda=0.5$ & 9.889 & {\bf 5.538} & 6.418 & 5.655 & 7.852 & 7.088 & 7.064 \\ \hline
$\lambda=0.6$ & 10.030 & 6.200 & 6.665 & {\bf 6.048} & 9.234 & 7.710 & 7.387 \\ \hline
$\lambda=0.7$ & 10.171 & 6.863 & 6.911 & {\bf 6.440} & 10.616 & 8.333 & 7.709 \\ \hline
$\lambda=0.8$ & 10.312 & 7.525 & 7.158 & {\bf 6.833} & 11.999 & 8.955 & 8.031 \\ \hline
$\lambda=0.9$ & 10.453 & 8.187 & 7.405 & {\bf 7.225} & 13.381 & 9.577 & 8.354 \\ \hline
$\lambda=1$ & 10.594 & 8.849 & 7.651 & {\bf 7.618} & 14.763 & 10.200 & 8.676 \\ \hline
\end{tabular}
\end{center}
\caption{$R_p$ values for $R=100\cdot\sum_{i=1}^{10} S_{(i)}/B$.}
\label{Table.Penalization2}
\end{table}

\begin{figure}[h]
\begin{center}
\begin{tabular}{@{}c@{}c@{}}
\includegraphics[width=.5\textwidth]{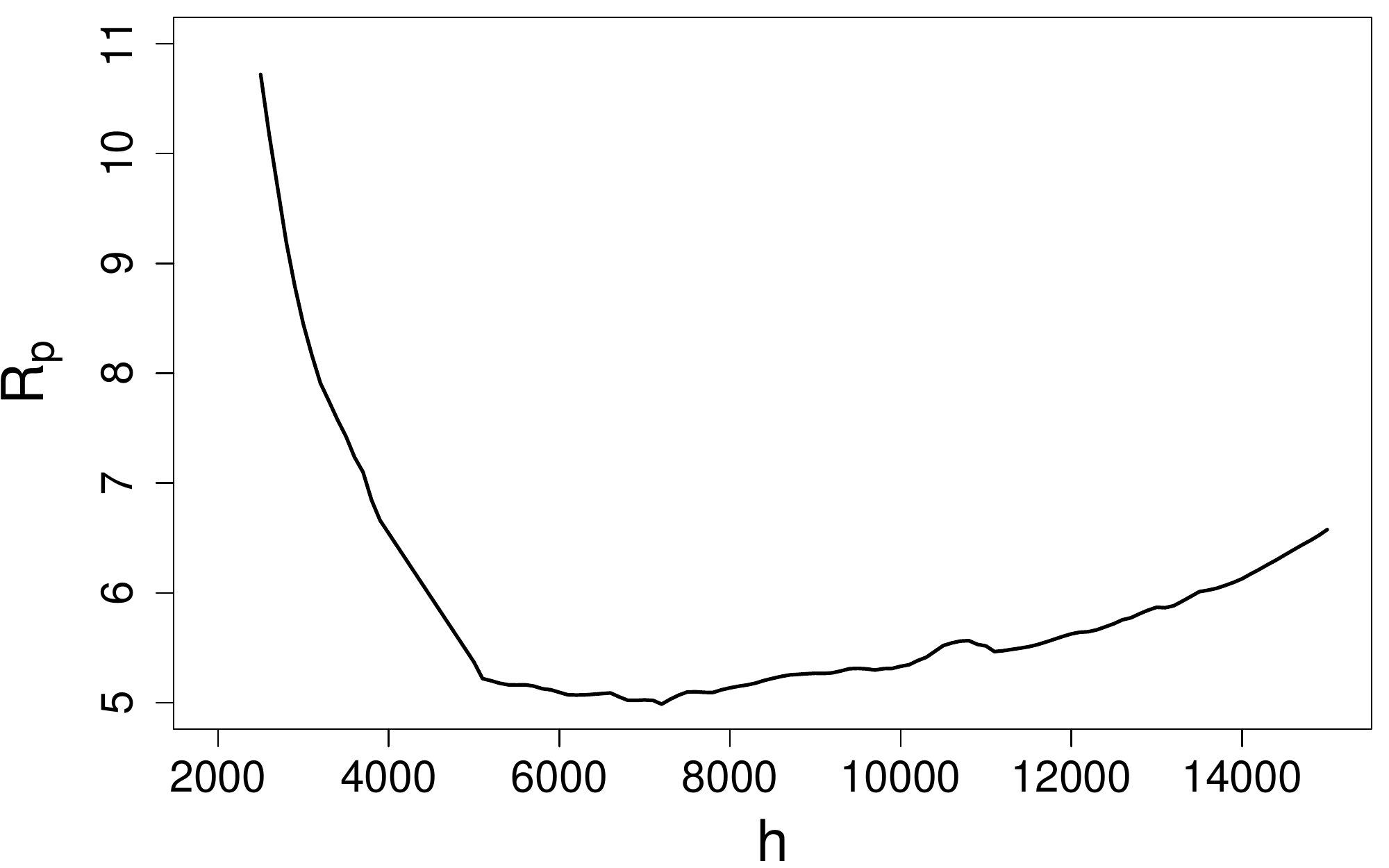} & \includegraphics[width=.5\textwidth]{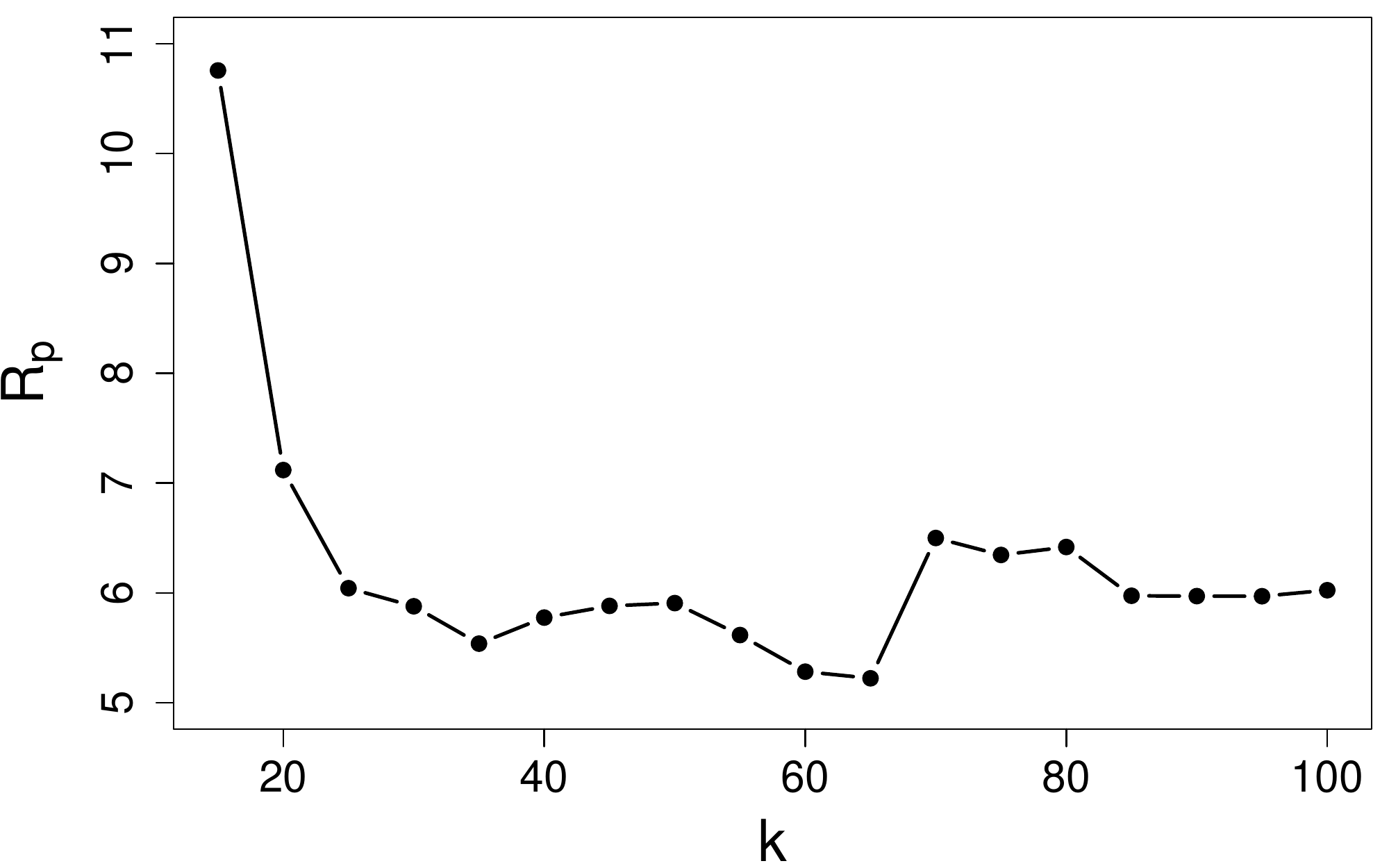}
\end{tabular}
\end{center}
\caption{$R_p$ as a function of the bandwidth for the kernel home range estimator (left) and as a function of the number of neighbours for the LoCoH estimator (right).}
\label{Figure.RpFunctions}
\end{figure}

Indeed, the new penalized criterion $R_p$ can be used as a tool for optimally choosing the parameters of the different home range estimators, by selecting the parameter value that minimises $R_p$. As an illustration, Figure \ref{Figure.RpFunctions} shows $R_p$ as a function of the bandwidth $h$ for the kernel home range estimator (left) and as a function of the number of neighbours $k$ for the LoCoH estimator (right), regarding the overestimation ratio $R=100\cdot \sum_{i=1}^{10}S_{(i)}/B$ and $\lambda=0.5$.

\begin{figure}[h]
\begin{center}
\includegraphics[width=9cm]{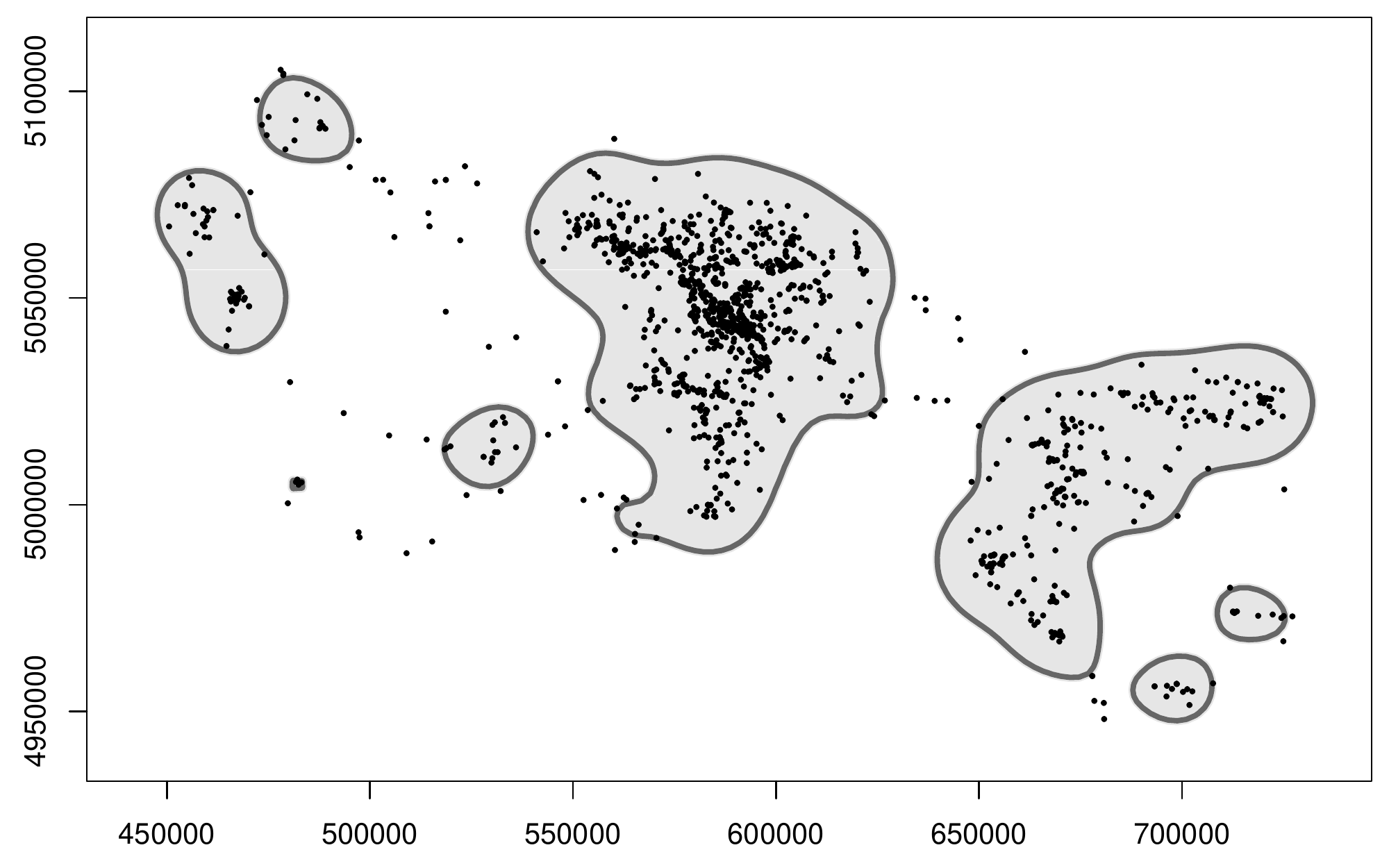}
\end{center}
\caption{Kernel home range estimator for the Zimzik locations using the optimal $h=7203.54$ for $\lambda=0.5$.}
\label{Figure.HR.opth}
\end{figure}

\begin{figure}[h]
\begin{center}
\includegraphics[width=9cm]{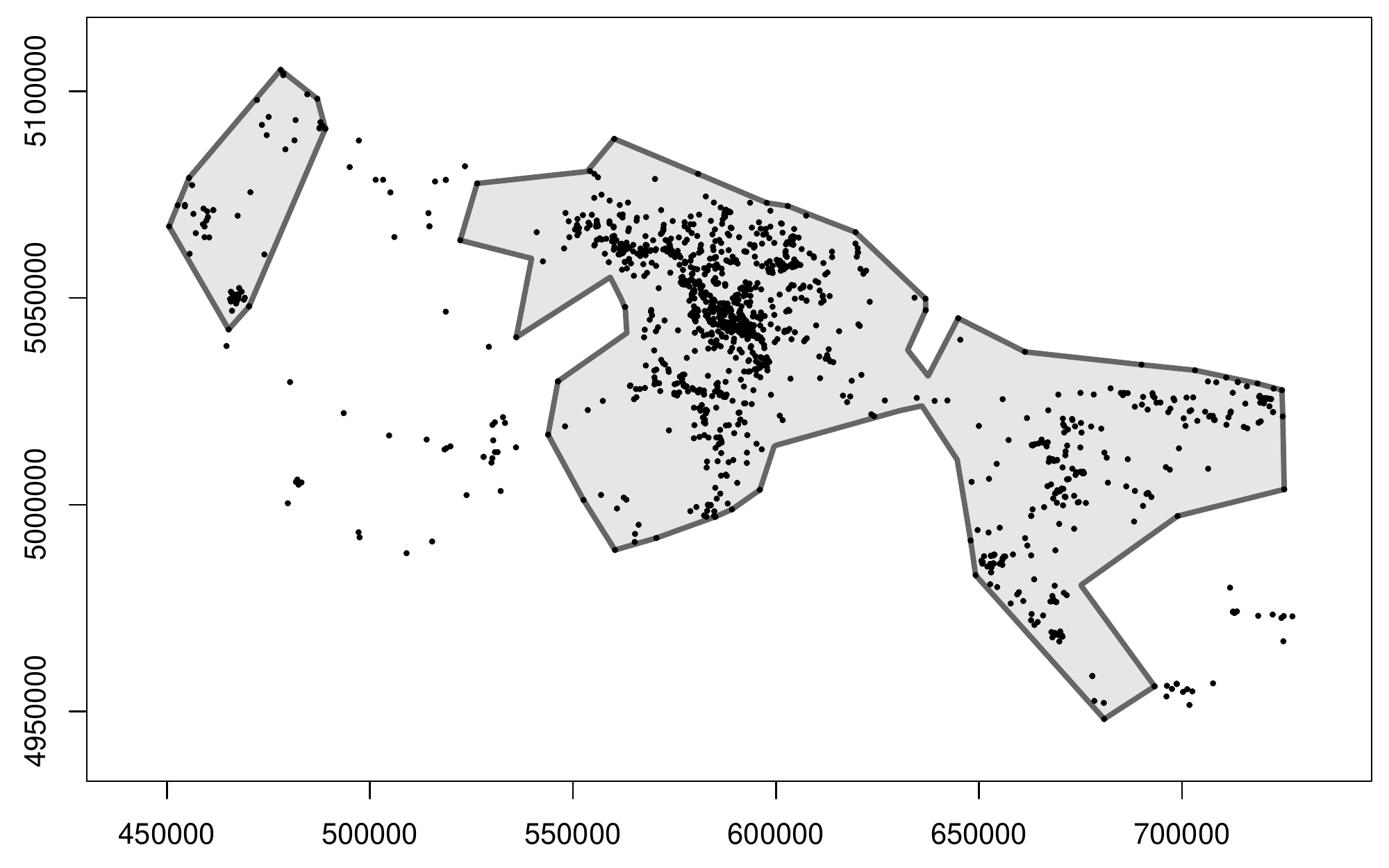}
\end{center}
\caption{LoCoH home range estimator for the Zimzik locations using the optimal value of $k=65$ for $\lambda=0.5$.}
\label{Figure.HR.optk}
\end{figure}

The bandwidth that minimises $R_p$ for the kernel home range estimator is $h=7203.54$, and the resulting $R_p$ value is $4.987$, which is lower than the value obtained by any other method in Table \ref{Table.Penalization2} (for $\lambda=0.5$). The corresponding home range, shown in Figure \ref{Figure.HR.opth}, clearly displays a compromise by presenting a small overestimation area while, at the same, avoiding sharp inlets into the location data shape. Analogously, the number of neighbours that minimises $R_p$ for the LoCoH estimator is $k=65$, resulting in an $R_p$ value of 5.223, which is also lower than for any other method in Table \ref{Table.Penalization2}, but still a bit higher than for the kernel estimator, meaning that in this situation the kernel home range attains a better balance between overestimation and too closely following the data locations. In any case, the LoCoH home range with this optimal value of $k$ is shown in Figure \ref{Figure.HR.optk}.

\section{Discussion}

In this work, as a first aim, we have reviewed all the proposals in
statistical home range estimation, up to the current state of the art.
Assuming that location data are independent, the best global estimation
procedure appears to be the kernel home range with an unconstrained
bandwidth matrix and the localized method performing best is the LoCoH.
Among the scarce methods taking into account the time dependence of the
observed locations, the most suitable ones are actually adaptations to
this context of the kernel and the LoCoH procedures, namely the MKDE and
the T-LoCoH respectively. These two procedures depend on tuning
parameters whose optimal choice has to be further investigated.

As a second objective, a natural consequence of this revision, we have
proposed a general procedure to select the most appropriate home range
among a collection of them based on the same set of relocations. The
selection is performed via an index measuring the excess extension of
the home range with respect to the relocations, penalized by the shape
circularity to prevent over-fitting to the sample. When computing this
penalization criterion on several home ranges based on the same real
data set, the kernel and the LoCoH home ranges are the ones optimizing
it. Further, optimization of the penalized selection index has led to a
good-performing, natural choice of the tuning parameters in these home
range estimators.

\section*{Acknowledgements}

{The authors are grateful to Antonio Cuevas for drawing their attention to the home range estimation problem and for some insightful discussions.}
The research on the Mongolian wolves data was conducted within the framework of the Przewalskii's horse reintroduction project of the International Takhi Group (ITG), in cooperation with the Mongolian Ministry of Nature and Environment, the National University in Ulaanbaatar, Mongolia and the Great Gobi B Strictly Protected Area Administration. Field work would not have been possible without the help of park director O. Ganbataar, former park director Suchebaatar, the rangers Batsuuri, Chinbat, Huder, Nisekhhuu, Enkhbaatar, Nyambayar and their families, nor without the help and support of the local people from Tachin Tal.
Funding for the research on the wolves data was provided by the Austrian Science Foundation (FWF) project P14992 and the Austrian National Bank (Jubil\"{a}ums Fonds) through the Zoo Salzburg (Research for Conservation).

\begin{center}
{\bf References}
\end{center}

\newenvironment{nonindentedreferences}
  {\begin{list}{}{\setlength{\labelwidth}{0pt}
   \setlength{\itemindent}{-5mm}
   \setlength{\leftmargin}{5mm}
   }}
  {\end{list}}

\begin{nonindentedreferences}\itemsep-1mm
\item Aaron, C. and Bodart, O. (2016). Local convex hull support and boundary estimation. {\em Journal of Multivariate Analysis}, 147, 82--101.
\item Bath, S.K., Hayter, A.J., Cairns, D.A. and Anderson, C. (2006). Characterization of home range using point peeling algorithms. {{\em Journal of Wildlife Management}, 70, 422--434.}
\item Bauder, J.M., Breininger, D.R., Bolt, M.R., Legare, M.L., Jenkins, C.L. and McGarigal, K. (2015). The role of the bandwidth matrix in influencing kernel home range estimates for snakes using VHF telemetry data. {\em Wildlife Research}, 42, 437--453.
\item Benhamou, S. and Cornelis, D. (2010). Incorporating movement behavior and barriers to improve biological relevance of kernel home range space use estimates. {\em Journal of Wildlife Management}, 74, 1353--1360.
\item Berger, K.M. and Gese, E.M. (2007). Does interference competition with wolves limit the distribution and abundance of coyotes? {\em Journal of Animal Ecology}, 76, 1075--1085.
\item Bertrand, M.R., DeNicola, A.J., Beissinger, S.R. and Swihart, R.K. (1996). Effects of parturition on home ranges and social affiliations of female white-tailed deer. {\em Journal of Wildlife Management}, 60, 899--909.
\item Blair, W.F. (1940). Notes on home ranges and populations of the short-tailed shrew. {\em Ecology}, 21, 284--288.
\item Burgman, M. A. and Fox, J. C. (2003). Bias in species range estimates from minimum convex polygons: implications for conservation and options for improved planning. {\em Animal Conservation}, 6, 19--28.
\item Burt, W.H. (1943). Territoriality and home range concepts as applied to mammals. {\em Journal of Mammalogy}, 24, 346--352.
\item Calenge, C. (2006). The package ``adehabitat'' for the R software: A tool for the analysis of space and habitat use by animals. {\em Ecological Modelling},  197, 516--519.
\item Calhoun, J.B. and Casby, J.U. (1958). Calculation of home range and density of small mammals. Public Health Monograph n. 55. US Department of Health, Education and Welfare. Washington, DC, USA.
\item Carey, A.B., Reid, J.A. and Horton, S.P. (1990). Spotted owl home range and habitat use in southern Oregon coast ranges. {\em Journal of Wildlife Management}, 54, 11--17.
\item Chac\'{o}n, J.E. and Duong, T. (2010). Multivariate plug-in bandwidth selection with unconstrained pilot bandwidth matrices. {\em Test}, 19, 375--398.
\item Cholaquidis, A., Cuevas, a. and Fraiman, R. (2014). On Poincar\'{e} cone property. {\em Annals of Statistics}, 42, 255--284.
\item Cholaquidis, A., Fraiman, R., Lugosi, G. and Pateiro-L\'{o}pez, B. (2016). Set estimation from reflected Brownian motion. {\em Journal of the Royal Statistical Society B}, 78, 1057--1078.
\item Cumming, G.S. and Corn\'{e}lis, D. (2012). Quantitative comparison and selection of home range metrics for telemetry data. {\em Diversity and Distributions}, 18, 1057--1065.
\item Devroye, L. and Krzy\.zak, A. (1999). On the Hilbert kernel density estimate. {\em Statistics and Probability Letters}, 44, 209--308.
\item Dixon, K.R. and Chapman, J.A. (1980). Harmonic mean measure of animal activity areas. {\em Ecology},  61, 1040--1044.
\item Don, B.A.C. and Rennolls, K. (1983). A home range model incorporating biological attraction points. {\em Journal of Animal Ecology}, 52, 69--81.
\item Dougherty, E.R., Carlson, C.J., Blackburn, J.K. and Getz, W.M. (2017). A cross-validation-based approach for delimiting reliable home range estimates. {\em Movement Ecology}, 5:19.
\item Downs, J.A. (2010). Time-geographic density estimation for moving point objects. In {\em Geographic Information Science}. Eds. Fabrikant, S.I., Reichenbacher, T., VanKreveld, M. and Schlieder, C. {\em Lecture Notes in Computer Science}, 6292, 16--26.
\item Downs, J.A. and Horner, M.W. (2009). A characteristic-hull based method for home range estimation. {\em Transactions in GIS}, 13, 527--537.
\item Downs, J.A., Horner, M.W. and Tucker, A.D. (2011). Time-geographic density estimation for home range analysis. {\em Annals of GIS}, 17, 163--171.
\item Duong, T. (2018). ks: Kernel Smoothing. R package version 1.11.0.
  https://CRAN.R-project.org/package=ks
\item Duong, T. and Hazelton, M.L. (2003). Plug-in bandwidth matrices for bivariate kernel density estimation. {\em Journal of Nonparametric Statistics}, 15, 17--30.
\item Edelsbrunner, H., Kirkpatrick, D. G. and Seidel, R. (1983). On the shape of a set of points in the plane. {\em IEEE Transactions on Information Theory}, 29, 551--559.
\item Fasy, B.T., Kim, J., Lecci, F., Maria, C. and Rouvreau, V. (2017). TDA: Statistical Tools for Topological Data Analysis. The included GUDHI is authored by Clement Maria, Dionysus by Dmitriy Morozov, PHAT by Ulrich Bauer, Michael Kerber and Jan Reininghaus. R package version 1.6. https://CRAN.R-project.org/package=TDA
\item Fleming, C.H., Fagan, W.F., Mueller, T., Olson, K.A., Leimgruber, P. and Calabrese, J.M. (2015). Rigorous home range estimation with movement data: a new autocorrelated kernel density estimator. {\em Ecology}, 96, 1182--1188.
\item Foley, A.M., Schroeder, B.A., Hardy, R., MacPherson, S.L. and Nicholas, M. (2014). Long-term behavior at foraging sites of adult female loggerhead sea turtles ({\em Caretta caretta}) from three Florida rookeries. {\em Marine Biology}, 161, 1251--1262.
\item Getz, W.M. and Wilmers, C.C. (2004). A local nearest-neighbor convex-hull construction of home ranges and utilization distributions. {\em Ecography}, 27, 489--505.
\item Getz, W.M., Fortmann-Roe, S., Cross, P.C., Lyons, A.J., Ryan, S.J. and Wilmers, C.C. (2007). LoCoH: nonparametric kernel methods for constructing home ranges and utilization distributions. {\em PLoS ONE}, 2, e207.
\item Girard, I., Ouellet, J.P., Courtois, R., Dussault, C. and Breton, L. (2002). Effects of sampling effort based on GPS telemetry on home-range size estimations. {\em Journal of Wildlife Management}, 66, 1290--1300.
\item Gonz\'{a}lez, R.C. and Woods, R.E. (2008). {\em Digital Image Processing}. Third edition. Pearson International Edition.
\item H\"{a}gerstrand, T. (1970).  What about people in regional science?. {\em Papers of the Regional Science Association}, 24, 6--21.
\item Hall, P., Lahiri, S.N. and Truong, Y.K. (1995) On bandwidth choice for density estimation with dependent data. {\em Annals of Statistics}, 23, 2241--2263.
\item Harris, S., Cresswell, W.J., Forde, P.G., Trewhella, W.J., Woollard, T., and Wray, S. (1990). Home-range analysis using radio-tracking data - A review of problems and techniques particularly as applied to the study of mammals. {\em Mammal Review}, 20, 97--123.
\item Harvey, M.J. and Barbour, R.W. (1965). Home range of {\em Microtus ochrogaster} as determined by a modified minimum area method. {\em Journal of Mammalogy}, 46, 398--402.
\item Hemson, G., Johnson, P., South, A., Kenward, R., Ripley, R. and Macdonald, D. (2005). Are kernels the mustard? Data from global positioning system (GPS) collars suggests problems for kernel homerange analyses with least-squares cross-validation. {\em Journal of Animal Ecology}, 74, 455--463.
\item Horne, J.S., Garton, E.O. and Rachlow, J.L. (2008). A synoptic model of animal space use: Simultaneous estimation of home range, habitat selection, and inter/intra-specific relationships. {\em Ecological Modelling}, 214, 338--348.
\item Jacques, C.N., Jenks, J.A. and Klaver, R.W. (2009). Seasonal movements and home-range use by female pronghorns in sagebrush-steppe communities of western South Dakota. {\em Journal of Mammalogy}, 90, 433--441.
\item Kaczensky, P., Ganbaatar, O., Enksaikhaan, N. and Walzer, C. (2006). Wolves in Great Gobi B SPA GPS tracking study 2003-2005 dataset. Movebank Data Repository (\url{www.movebank.org}).
\item Kaczensky, P., Enkhsaikhan, N., Ganbaatar, O. and Walzer, C. (2008). The Great Gobi B Strictly Protected Area in Mongolia - refuge or sink for wolves Canis lupus in the Gobi. {\em Wildlife Biology}, 14, 444--456.
\item Keating, K.A. and Cherry, S. (2009). Modeling utilization distributions in space and time. {\em Ecology}, 90, 1971--1980.
\item Kenward, R.E., Clarke, R.T., Hodder, K.H., and Walls, S.S. (2001). Density and linkage estimators of home range: nearest-neighbor clustering defines multinuclear cores. {\em Ecology}, 82, 1905--1920.
\item Kie, J.G., and Boroski, B.B. (1996). Cattle distribution, habitats, and diets in the Sierra Nevada of California. {\em Journal of Range Management}, 49, 482--488.
\item Kie, J.G., Matthiopoulos, J., Fieberg, J., Powell, R.A., Cagnacci, F., Mitchell, M.S., Gaillard, J.-M. and Moorcroft, P.R. (2010). The home-range concept: are traditional estimators still relevant with modern telemetry technology?. {\em Philosophical Transactions of the Royal Society B}, 365, 2221--2231.
\item Kranstauber, B., Kays, R., LaPoint, S.D., Wikelski, M. and Safi, K. (2012). A dynamic Brownian bridge movement model to estimate utilization distributions for heterogeneous animal movement. {\em Journal of Animal Ecology}, 81, 738--746.
\item List, R. and Macdonald B.W. (2003). Home range and habitat use of the kit fox ({\em Vulpes macrotis}) in a prairie dog ({\em Cynomys ludovicianus}) complex. {\em Journal of Zoology}, 259, 1--5.
\item Long, J. (2017). wildlifeTG: Time Geograhic Analysis of Wildlife Telemetry Data. R package
  version 0.4. \url{http://jedalong.github.io/wildlifeTG}
\item Long, J. and Nelson, T. (2012). Time geography and wildlife home range delineation. {\em The Journal of Wildlife Management}, 76, 407--413.
\item Long, J. and Nelson, T. (2015). Home range and habitat analysis using dynamic time geography. {\em The Journal of Wildlife Management}, 79, 481--490.
\item Lyons, A., Getz, W. and the R Development Core Team (2018). T-LoCoH: Time Local Convex
  Hull Homerange and Time Use Analysis. R package version 1.40.05.
\item Lyons, A.J., Turner, W.C. and Getz, W.M. (2013). Home range plus: a space-time characterization of movement over real landscapes. {\em Movement Ecology}, 1:2.
\item Matthiopoulos, J. (2003). Model-supervised kernel smoothing for the estimation of spatial usage. {\em Oikos}, 102, 367--377.
\item Meekan, M.G., Duarte, C.M., Fern\'{a}ndez-Gracia, J., Thums, M., Sequeira, A.M.M., Harcourt, R. and Eguiluz, V.M. (2017).  The ecology of human mobility. {\em Trends in Ecology \& Evolution}, 32, 198--210.
\item Miller, H.J. (2005). A measurement theory for time geography. {\em Geographical Analysis}, 37, 17--45.
\item Mohr, C.O. (1947). Table of equivalent populations of North American small mammals. {\em American Midland Naturalist}, 37, 223--449.
\item Moorcroft, P.R. and Lewis, M.A. (2006). {\em Mechanistic Home Range Analysis}. Princeton University Press.
\item Nilsen, E.B. Pedersen, S. and Linnell, J.D.C. (2008). Can minimum convex polygon home ranges be used to draw biologically meaningful conclusions?. {\em Ecological Research}, 23, 635--639.
\item Odum, E.P. and Kuenzler, E. (1955). Measurement of territory and home range size in birds. {\em Auk}, 72, 128--137.
\item Pateiro-L\'{o}pez, B. and Rodr\'{\i}guez-Casal, A. (2010). Generalizing the convex hull of a sample: the R package alphahull. {\em Journal of Statistical Software}, 34, 1--28.
\item Pateiro-L\'{o}pez, B. and Rodr\'{\i}guez-Casal, A. (2016). alphahull: Generalization of the Convex Hull of a Sample of Points in the Plane. R package version 2.1.
\item Pavey, C.R., Goodship, N and Geiser, F. (2003). Home range and spatial organisation of rock-dwelling carnivorous marsupial, {\em Pseudantechinus macdonnellensis}. {\em Wildlife Research}, 30, 135--142.
\item Pellerin, M., Sa\"{\i}d, S. and Gaillard, M. (2008). Roe deer {\em Capreolus capreolus} home-range sizes estimated from VHF and GPS data. {\em Wildlife Biology}, 14, 101--110.
\item Perkal, J. (1956). Sur les ensembles $\varepsilon$-convexes. {\em Colloquium Mathematicae}, 4, 1--10.
\item R Core Team (2018). R: A language and environment for statistical computing. R Foundation for
  Statistical Computing, Vienna, Austria. URL \url{www.R-project.org/}.
\item Rodr\'{\i}guez-Casal, A. (2007). Set estimation under convexity type assumptions. {\em Annales de l'I.H.P.- Probabilit\'{e}s \& Statistiques}, 43, 763--774.
\item Rodr\'{\i}guez-Casal, A. and Saavedra-Nieves, P. (2016).  A fully data-driven method for estimating the shape of a point cloud. {\em ESAIM: Probability and Statistics}, 20, 332--348.
\item Seaman, D.E., Millspaugh, J.J., Kernohan, B.J., Brundige, G.C., Raedeke, K.J. and Gitzen, R.A. (1999). Effects of sample size on kernel home range estimates. {\em Journal of Wildlife Management}, 63, 739--747.
\item Seaman, D.E. and Powell, R.A. (1996). An evaluation of the accuracy of kernel density estimators for home range analysis. {\em Ecology}, 77, 2075--2085.
\item Seton, E.T. (1909). {\em Life-histories of Northern Animals: an Account of the Mammals of Manitoba. Vol. I-Grass-Eaters}. Charles Scribner's Sons.
\item Signer, J., Balkenhol, N., Ditmer, M. and Fieberg, J. (2015). {\em Animal Biotelemetry}, 3, 16.
\item Steiniger, S. and Hunter, A.J.S. (2013). A scaled line-based density estimator for the retrieval of utilization distributions and home ranges from GPS movement tracks. {\em Ecological Informatics}, 13, 1--8.
\item Tarjan, L.M. and Tinker, M.T. (2016). Permissible home range estimation (PHRE) in restricted habitats: a new algorithm and an evaluation for sea otters. {\em PLoS One}, 11(3), e0150547.
\item Tracey, J.A., Sheppard, J., Zhu, J., Wei, F., Swaisgood, R.R. and Fisher, R.N.  (2014). Movement-based estimation and visualization of space use in 3D for wildlife ecology and conservation. {\em PLOS One}, 9, e101205.
\item Walsh, P.D., Boyer, D. and Crofoot, M.C. (2010). Monkey and cell-phone-user mobilities scale similarly. {\em Nature Physics}, 6, 929--930.
\item Walter, W.D., Onorato, D.P. and Fischer, J.W. (2015). Is there a single best estimator? Selection of home range estimators using area-under-the-curve. {\em Movement Ecology}, 3:10.
\item Wasserman, L. (2018). Topological data analysis. {\it Annual Review of Statistics and Its Application}, 5, 501--532.
\item van Winkle, W. (1975). Comparison of several probabilistic home-range models. {\em The Journal of Wildlife Management}, 39, 118--123.
\item Worton, B.J. (1987). A review of models of home range for animal movement. {\em Ecological Modelling}, 38, 277--298.
\item Worton, B.J. (1989). Kernel methods for estimating the utilization distribution in home-range studies. {\em Ecology}, 70, 164--168.
\end{nonindentedreferences}

\end{document}